\documentclass[12pt]{article}
\usepackage[utf8]{inputenc}
\usepackage[left]{lineno}
\usepackage{setspace}

\usepackage{latexsym,amsmath,amssymb,amsfonts,amsthm}
\usepackage{bm} 

\usepackage{graphicx}
\usepackage{booktabs}
\usepackage{multirow}
\usepackage{adjustbox}
\usepackage{subcaption}
\usepackage{mathrsfs}
\usepackage{mathabx}

\usepackage{natbib}
\usepackage[colorlinks,citecolor=blue,urlcolor=blue]{hyperref}

\usepackage{setspace}
\usepackage{appendix}
\usepackage{enumitem}
\usepackage{authblk}
\usepackage{tikz}
\usepackage{float}
\usepackage{makecell}
\usepackage{bbm} 
\usepackage{lscape} 
\usepackage{longtable}
\usepackage{etoolbox}
\usepackage{listofitems}

\usepackage{xcolor}
\usepackage{url}
\usepackage{bbm}
\usepackage{lmodern} 
\usepackage[breakable,skins]{tcolorbox} 
\usepackage{textcomp} 
\usepackage{wrapfig} 
\usepackage[english]{babel}
\usepackage{amsfonts}
\usepackage{amssymb}
\usepackage{graphicx}
\usepackage{enumerate}
\usepackage{setspace}
\theoremstyle{definition}

\allowdisplaybreaks

\definecolor{lb}{RGB}{44,139,183}

\newcommand{\mbf}{\mathbf}
\newcommand{\bfs}{\mathbf{s}}
\newcommand{\bfA}{\mathbf{A}}
\newcommand{\bfb}{\mathbf{b}}
\newcommand{\bfx}{\mathbf{x}}

\newcommand{\bfv}{\mathbf{v}}
\newcommand{\bfu}{\mathbf{u}}

\newcommand{\bfG}{\mathbf{G}}
\newcommand{\bfY}{\mathbf{Y}}

\newcommand{\bms}{\boldsymbol}
\newcommand{\bfgamma}{\bms{\gamma}}
\newcommand{\bfkappa}{\bms{\kappa}}
\newcommand{\bftheta}{\bms{\theta}}
\newcommand{\bfphi}{\bms{\phi}}

\newcommand{\bfomega}{\bms{\omega}}
\newcommand{\bfeta}{\bms{\eta}}
\newcommand{\bfSigma}{\bms{\Sigma}}

\newcommand{\intd}{\mathrm{d}}

\def\frac#1#2{{\textstyle{#1\over#2}}}

\DeclareSymbolFont{AMSb}{U}{msb}{m}{n}
\DeclareMathSymbol{\Natural}{\mathbin}{AMSb}{"4E}
\DeclareMathSymbol{\Integer}{\mathbin}{AMSb}{"5A}
\DeclareMathSymbol{\Real}{\mathbin}{AMSb}{"52}
\DeclareMathSymbol{\Rational}{\mathbin}{AMSb}{"51}
\DeclareMathSymbol{\Imaginary}{\mathbin}{AMSb}{"49}
\DeclareMathSymbol{\Complex}{\mathbin}{AMSb}{"43} 
\DeclareMathSymbol{\Disk}{\mathbin}{AMSb}{"44} 
\def\bi{\begin{itemize}}
\def\ei{\end{itemize}}
\def\bd{\begin{description}}
\def\ed{\end{description}}
\def\ben{\begin{enumerate}}
\def\een{\end{enumerate}}
\def\bv{\begin{verbatim}}
\def\ev{\end{verbatim}}

\def\hat#1{{\widehat{#1}}}

\def\2to{{\ {\buildrel 2\over \longrightarrow}\ }}

\def\I1ton{{$I_1,\ldots,I_n$}}
\def\X1ton{{$X_1,\ldots,X_n$}}
\def\Y1ton{{$Y_1,\ldots,Y_n$}}
\def\Z1ton{{$Z_1,\ldots,Z_n$}}
\def\R1ton{{$R_1,\ldots,R_n$}}
\def\e1ton{{$e_1,\ldots,e_n$}}
\def\t1ton{{$t_1,\ldots,t_n$}}
\def\x1ton{{$x_1,\ldots,x_n$}}
\def\y1ton{{$y_1,\ldots,y_n$}}
\def\z1ton{{$z_1,\ldots,z_n$}}

\title{Spatio-temporal modeling and forecasting with Fourier neural operators}

\author[1]{Pratik Nag}
\author[1]{Andrew Zammit-Mangion}
\author[1]{Sumeetpal Singh}
\author[1]{Noel Cressie}

\affil[1]{School of Mathematics and Applied Statistics, University of Wollongong, Australia}
\date{}
\begin{document}


\maketitle

\begin{abstract}
Spatio-temporal process models are often used for modeling dynamic physical and biological phenomena that evolve across space and time. These phenomena may exhibit environmental heterogeneity
and complex interactions that are difficult to capture using traditional
statistical process models such as Gaussian processes. This
work proposes the use of Fourier neural operators (FNOs) for constructing
statistical dynamical spatio-temporal models for forecasting.
An FNO is a flexible mapping of functions that approximates the solution operator of possibly unknown linear or non-linear partial differential equations (PDEs) in a computationally efficient manner. It does so using samples of inputs and their respective outputs, and hence explicit knowledge of the underlying PDE is not required. Through simulations from a nonlinear PDE with known solution, we compare FNO forecasts to those from state-of-the-art statistical spatio-temporal-forecasting methods. Further, using sea surface temperature
data over the Atlantic Ocean and precipitation data across Europe,
we demonstrate the ability of FNO-based dynamic spatio-temporal (DST) statistical modeling to capture complex
real-world spatio-temporal dependencies. Using collections of testing instances, we show that the FNO-DST forecasts are accurate with valid uncertainty quantification.

\end{abstract}


\baselineskip=26pt

\section{Introduction}

Spatio-temporal (ST) modeling and data analysis is often a component of physical, environmental, and biological studies. Such ST processes can exhibit complex dependencies across both space and time and hence can be difficult to model. ST statistical models fall into two main categories \citep{wikle2019spatio}. The first category consists of \textit{descriptive} ST models, such as Gaussian processes \citep[GPs; e.g., ][]{cressie1999classes, gneiting2007,  cressie2011statistics,banerjee2014hierarchical}, where the time dimension is simply considered to be an extra dimension and treated in the same way as the spatial dimensions. This approach often faces challenges in computational scalability and its handling of complex dynamics. Moreover, these descriptive ST models typically rely on predefined classes of spatio-temporal covariance functions, which limit their flexibility.

The second category consists of \textit{dynamical} ST (DST) models, which are constructed by specifying conditional-dependence relationships between and within spatial fields at successive time steps \citep[e.g.,][Ch.7]{cressie2011statistics}. These models provide an elegant framework for capturing the temporal evolution of a spatial process, making them particularly effective in applications where the underlying (physical, biological, etc.) system dynamics are complex. DST models offer a distinct advantage over descriptive ST models since they are able to encode the scientific process being modeled, which makes them well suited for forecasting tasks.

A class of DST models that has been extensively used for modeling spatio-temporal dynamical systems is the class of integro-difference equation (IDE) models \citep{wikle1999, wikle2002}. These models are designed to characterize the conditional dependence between a spatial field at a future time step and its current state, through an integral operator. They have been successfully applied in various domains, such as ecology \citep[e.g.,][]{kot1996}, meteorology \citep[e.g.,][]{xu2005,calder2011}, and epidemiology \citep[e.g.,][]{kot1986}. 

Typically, DST models assume a linear operator, which simplifies their mathematical formulation. However, the linearity assumption may not be appropriate over long time horizons, where environmental heterogeneity and dynamic interactions can induce strong nonlinear dependencies. One way to address this is through quadratic nonlinear IDE-based DST models \citep{wikle2010general,wikle2011}. While these models are better at capturing complex nonstationary behaviors, they are less computationally efficient for making statistical inferences. 

To mitigate the computational cost of fitting flexible DST models, \citet{de_bezenac_2019} proposed the use of convolutional neural networks (CNNs) for learning dynamical parameters from data. Their approach was extended to stochastic IDE models by \citet{ZAMMITMANGION2020100408}. In both of these works, the CNN is trained offline but, once trained, parameter estimation, followed by forecasting, can be done with very little computational cost.
The framework relies on a predefined governing linear partial differential equation (PDE), the solution to which is used to construct the IDE. However, many real-world applications involve heterogeneous or highly nonlinear dynamics that are not well represented through linear PDEs.

Recent advances in deep learning have revolutionized researchers' ability to model complex systems, particularly those governed by PDEs, linear or nonlinear. One such advance is the Fourier neural operator \citep[FNO;][]{li2020neural}, which are building blocks for modeling intricate physical dynamics of spatio-temporal processes. Unlike traditional solvers that explicitly solve PDEs, FNOs use training data to learn mappings between functional spaces based on samples of inputs and their respective outputs. This makes them particularly suited for spatio-temporal-forecasting applications where the underlying dynamics may not be known. Notably, they are more computationally efficient than standard PDE solvers, which makes them an attractive alternative to stochastic physical models. 

In this study, we introduce a general class of stochastic DST models that uses an FNO framework to capture the temporal evolution of spatial processes; we refer to these as \textit{FNO-DST models}. Their forecasting performance is evaluated here through simulation studies involving Burgers' equation, a nonlinear (non-stochastic) PDE whose forecasts can be obtained numerically through finite-difference approximations. Then we consider forecasting of geophysical variables. We use FNO-DSTs to forecast sea surface temperature (SST) over the North Atlantic Ocean, where it has been shown that the dynamics can be well represented through an advection-diffusion PDE \citep{de_bezenac_2019}. In a second application, we use FNO-DSTs to forecast precipitation in a region over Europe, where the physical dynamics are complex and unknown. In both applications, we find that the FNO-DST model captures complex spatio-temporal interactions and that this leads to accurate forecasts.

The remainder of the paper is structured as follows: Section \ref{sec:methodology} gives the modeling and methodological details of an FNO, and it describes how we incorporate it into an FNO-DST model. Section \ref{sec:experiments} details the setup of the simulation experiment that compares the performance of the FNO-DST forecasts to other spatio-temporal forecasts, where the spatio-temporal data are generated using Burgers' equation that recall is nonlinear. Section \ref{sec:application} applies the FNO-DST approach to two geophysical variables, and we again compare its forecasting performance to that of other approaches. Section \ref{sec:conclusion} discusses key findings and potential directions for future research. This article is accompanied by on-line Supplementary Material.

\section{The FNO-DST model}\label{sec:methodology}

In Section \ref{subsec:IDE}, we describe IDE models and a nonlinear generalization that employs neural operators. In Section \ref{subsec:FNO}, we introduce the FNO-DST model as a computationally tractable specification of the neural-operator-based model discussed in Section \ref{subsec:IDE}. Parameter estimation, specifically maximum likelihood estimation of the FNO-DST model's parameters, is presented in Section \ref{subsec:Inference}.

\subsection{Classical IDE-based dynamical spatio-temporal models}\label{subsec:IDE}
 Let $\mathcal{T} = [0,K]$ denote a continuous-time domain,  and let $\mathcal{D} \subset \mathbb{R}^d$ be a spatial domain; in our examples, $d$ is either 1 or 2. Although many models for environmental processes of interest assume continuous evolution on $\mathcal{D} \times \mathcal{T}$, for computational expediency their solutions are often discretized, resulting in a correlated time series of spatial processes. We adopt the same approach here: Let \(\{\, t_1, t_2, \dots, t_{T} \,\}\) be a finite set of time points in $\mathcal{T}$ with fixed increments $t_{k+1} - t_k = \delta > 0$, where $\delta \ll K$, and where $0\leq t_1 \leq t_T \leq K$. For spatio-temporal processes governed by PDEs, a discrete-time representation can sometimes be constructed from integral solutions at these successive time points.
 As an example, we choose a simple linear advection-diffusion PDE, here a real-valued continuous-time spatial process $\mathcal{C}(\bfs,t)$ on $\mathcal{D} \times [t_k , t_{k+h}]$, that evolves according to, 
 \begin{equation}
\dfrac{\partial \mathcal{C}(\bfs,t)}{\partial t} + \gamma_1 \sum_{i=1}^{d} \dfrac{\partial  \mathcal{C}(\bfs,t)}{\partial s_i} - \gamma_2 \sum_{i=1}^{d}  \dfrac{\partial^2 \mathcal{C}(\bfs,t)}{\partial s_i^2} 
     = 0; \quad \bfs \in \mathcal{D},~~t \in [t_k,t_{k+h}], 
\label{eq:advection-diffusion}
\end{equation}
where the vector of parameters $\bfgamma \equiv (\gamma_1,\gamma_2)^\top \in \mathbb{R} \times \mathbb{R}^{+} \equiv \Gamma$, and where $h$ is a fixed positive integer such that the end-point $t_{k+h} \in \mathcal{T}$. For fixed and known $\bfgamma$, $\mathcal{D}$ bounded, and with suitable regularity conditions \citep{brezis2011functional,evans2010pde} on $\mathcal{C}(\bfs, t_k)$, the solution to \eqref{eq:advection-diffusion} at the end-point $t_{k+h}$ involves a Green's function $g(\cdot, \cdot)$, and is \citep{brauer1967green,Pan2009}:
\begin{equation}
    \mathcal{C}(\bfs,t_{k+h}) = \mathcal{G}_{\bfgamma}[\mathcal{C}(\cdot,t_{k})]  \equiv \int_{\mathcal{D}} g(\bfs - \mbf{u},h\delta;\bfgamma)\mathcal{C}(\bfu,t_{k}) \textrm{d}\mbf{u};\quad \bfs \in \mathcal{D}, k = 1,2,\ldots, T-h. \label{eq:advection-solution}
\end{equation}
 In \eqref{eq:advection-solution}, we have used $\mathcal{G}_{\bfgamma}$ to denote the integral operator that maps the input function $\mathcal{C}(\cdot,t_{k}) \in L^2(\mathcal{D}; \mathbb{R})$ to a corresponding output function in the same space $L^2(\mathcal{D}; \mathbb{R})$, where $L^2(\mathcal{D}; \mathbb{R})$ is the space of squared integrable functions that map from $\mathcal{D} $ to $\mathbb{R}$. In \eqref{eq:advection-solution}, the Green's function \( g(\mathbf{s}-\mathbf{u},h\delta;\boldsymbol{\gamma})
= (4\pi\gamma_2 h\delta)^{-d/2}
\exp\!\big[-\|\mathbf{s}-\mathbf{u}-\gamma_1 h\delta\,\mathbf{1}\|^2 / (4\gamma_2 h\delta)\big] \), which is derived in the Supplementary Material, Section S.1.
The integral solution \eqref{eq:advection-solution} is a deterministic linear autoregression of the spatial process with time step $h\delta$ and starting spatial field $\mathcal{C}(\bfs,t_{k}) $.

We now introduce two sources of uncertainty that change \eqref{eq:advection-solution} into a stochastic process. First, we add a spatially correlated random term $\eta(\bfs,t_{k+h}), \bfs \in \mathcal{D}$, to \eqref{eq:advection-solution}, which captures variability of the underlying phenomenon being modeled and not accounted for by the integral solution. In this work, the spatial processes $\eta(\cdot, t_{k+h})$ and $\eta(\cdot, t_{k+h+{k'}})$ are assumed to be uncorrelated when conditioned on $Y_k(\cdot,t_k)$, for $k > 0$ and $k' < 0$ (their precise probabilistic dependence across time is made clear subsequently in \eqref{eq:cov_function}). Further, the processes $\{\eta(\bfs,t_k)\}_k$ are also each assumed to be at least second-order differentiable with respect to the spatial index $\bfs$, as is the solution to \eqref{eq:advection-diffusion}. Second, we model the parameters $\bfgamma$ in \eqref{eq:advection-solution} as random and independent of the process $\eta(\cdot, \cdot)$ entirely, across both space and time. These modeling assumptions lead to the following integro-difference equation model,
\begin{equation}\label{eq:IDE}
Y_{k+h}(\bfs) = \int_{\mathcal{D}} g(\bfs - \mbf{u}, h\delta;\bfgamma)Y_k(\bfu)\textrm{d}\bfu + \eta(\bfs, t_{k+h}); \quad \bfs \in \mathcal{D}, \ k = 1,2, \dots, T-h,
\end{equation}
where $g(\cdot, \cdot)$ is the Green's function, $Y_k(\cdot) \equiv \mathcal{C}(\cdot, t_k)$ for $k = 1,\dots,T-h$, and $Y_k(\cdot) \equiv Y(\cdot, t_k)$ is the stochastic process at time point $t_k$. Note that,
conditional on $\bfgamma$, the $h$-step processes defined in \eqref{eq:IDE} are Markov in time. For example when $h=1$, conditional on $\bfgamma$, the process {$\{Y_k(\cdot)\}_k$ is Markov in time. However, we shall see that, marginally, the past and the future of the process are not independent conditional on the present. 

Typically, a finite-dimensional representation of $Y_{k+h}(\cdot)$ is used. Here, we discretize $\mathcal{D}$ into basic areal units (BAUs) of common area $\Delta$. Let $\mathcal{D}^G \subset \mathcal{D}$ denote the BAU centroids, and define the random column vectors ${\bfY_k \equiv (Y_k(\bfs): \bfs \in \mathcal{D}^G)^\top}$ and $\bfeta_{k+h} \equiv (\eta(\bfs,t_{k+h}): \bfs \in \mathcal{D}^G)^\top$.
Approximating the integral in \eqref{eq:IDE} with a sum, we see that \eqref{eq:IDE} can be represented as the $|\mathcal{D}^G|$-dimensional vector autoregressive (VAR) process, 
\begin{equation}
      \bfY_{k+h} = \bfG_{\bfgamma} \bfY_k + \bfeta_{k+h};\quad k = 1,2,\dots,T-h,
    \label{eq:advection-solution-disc}
  \end{equation}
  where $\bfG_{\bfgamma} \equiv (g(\bfs - \mbf{u},h\delta;\bfgamma)\Delta: \bfs, \bfu \in \mathcal{D}^G)$ is a $|\mathcal{D}^G| \times |\mathcal{D}^G|$ matrix, and recall that $\bfgamma$ and hence $\bfG_{\bfgamma}$ are random. Now suppose that $\bfeta_{k+h}$ is a mean-zero $|\mathcal{D}^G|$-dimensional Gaussian vector such that $\textrm{var}(\bfeta_{k+h} \mid \bfY_k) = \bfSigma_{k+h}(\bfY_k; \bms{\alpha})$ with parameters $\bms{\alpha}$. Furthermore, assume that $\textrm{cov}(\bms{\eta}_k, \bfeta_{k+h} \mid \bfY_k) = \bms{0}$ for $k = (h+1),\dots, (T-h)$. Then,
\begin{equation}
    \bfY_{k+h} \mid \bfY_{k},\bfgamma  \sim \text{Gau}\left(\bfG_{\bfgamma}\bfY_k, \bfSigma_{k+h}(\bfY_k; \bms{\alpha})\right);\quad k = 1,2,\dots, T-h.
    \label{eq:IDE_dist}
\end{equation}

In this work, we are mainly interested in forecasting the process $\bfY_{k+h}$ at $t_{k+h}$ from information up to and including $t_k$. Now, the forecasting distribution given only information $\bfY_{k}$ at $t_k$ is
\begin{equation}
    p(\bfY_{k+h} \mid \bfY_{k}) = \int_{\Gamma} p(\bfY_{k+h} \mid \mbf{Y}_{k},\bfgamma) p(\bfgamma \mid \bfY_k)\textrm{d}\bfgamma.
\label{eq:IDE_dist-rgamma}
\end{equation}
Forecasting based on the predictive distribution given by \eqref{eq:IDE_dist-rgamma} is likely to be of little value since there is scarce information on the dynamical parameters $\bfgamma$ in just \(\bfY_k\) at time point \(k\). That is, $p(\bfgamma \mid \bfY_k) \approx p(\bfgamma)$ in \eqref{eq:IDE_dist-rgamma}. 

Intuitively, the dynamical parameters could be learned by incorporating information from recent history up to the time lag $\tau$: 
\begin{align}
  p(\bfY_{k+h} \mid \bfY_{k},\dots,\bfY_{k-\tau}) &= \int_{\Gamma} p(\bfY_{k+h} \mid \mbf{Y}_{k},\bfgamma) p(\bfgamma \mid \bfY_{k},\dots,\bfY_{k-\tau})\textrm{d}\bfgamma,
\label{eq:IDE_dist-rgamma-history}
\end{align}
where the conditional distribution of $\bfgamma$ is now used to sharpen the forecasting distribution. Now, as $\tau $ increases, the conditional distribution of $\bfgamma$ can be expected to concentrate its probability around the true parameter, which we denote as $\bfgamma^o$; that is, from \eqref{eq:IDE_dist-rgamma-history}, ${p(\bfY_{k+h} \mid \bfY_{k},\dots,\bfY_{k-\tau})} \approx p(\bfY_{k+h} \mid \bfY_{k}, \bfgamma^o)$ for large $\tau$. This brief discussion reveals that when developing spatio-temporal forecasting models in the absence of knowledge on the governing dynamical parameters, using a value of $\tau > 0$ that includes the past, is essential, even when the latent spatio-temporal model is conditionally Markov across time.

There are several ways to incorporate this history dependence into a temporally dynamic spatio-temporal model with hidden dynamical parameters. One approach is to replace the spatial convolution in \eqref{eq:IDE} with a sum (over time) of $(\tau+1)$  spatial convolutions, resulting in,
\begin{equation}\label{eq:IDE_history}
Y_{k+h}(\bfs) = \sum_{j=0}^\tau \int_\mathcal{D} g_{j}(\bfs - \mbf{u}, h\delta;\bftheta')Y_{k-j}(\bfu)\textrm{d}\bfu + \eta(\bfs, t_{k+h}); \quad \bfs \in \mathcal{D}, \quad k = \tau+1, \ldots, T-h,
\end{equation}
where $\{g_{j}(\cdot, h\delta;\bftheta')\}_{j = 0}^\tau$ are kernels parameterized by $\bftheta'$. Equation \eqref{eq:IDE_history} reduces to \eqref{eq:IDE} when $\tau=0$ and $g_0(\cdot, \cdot ) = g(\cdot, \cdot)$. As we did to obtain \eqref{eq:advection-solution-disc}, we can discretize $\mathcal{D}$ into $\mathcal{D}^G$ to obtain a VAR process,
\begin{equation}\label{eq:IDE2}
\bfY_{k+h} = \sum_{j=0}^\tau \bfG_{j,\bftheta'} \bfY_{k-j} + \bfeta_{k+h}; \quad k = \tau+1,\ldots, T-h,
\end{equation}
where $\bfG_{j,\bftheta'} \equiv (g_j(\bfs - \bfu, h\delta;\bftheta')\Delta: \bfs, \bfu \in \mathcal{D}^G)$ is a $|\mathcal{D}^G| \times |\mathcal{D}^G|$ matrix.

Another approach to incorporating history dependence, which relaxes the linearity structure of \eqref{eq:IDE_history}, is to make the kernel in \eqref{eq:IDE} state-dependent; this leads to a state-dependent VAR process (\citealp[][Sec.~7.3.2]{cressie2011statistics}, \citealp{ZAMMITMANGION2020100408}) of the form,
\begin{equation}\label{eq:IDE3}
\bfY_{k+h} = \bfG_{\bftheta''}(\bfY_{k},\dots,\bfY_{k-\tau}) \bfY_{k} + \bfeta_{k+h}; \quad k = \tau+1,\ldots, T-h.
\end{equation}
While both of these approaches lead to flexible classes of spatio-temporal models, practical implementations involve structure and restrictions on the matrices in \eqref{eq:IDE2} and \eqref{eq:IDE3}, and constant regular gridding. Fitting these models to data is also computationally expensive. Fourier neural operator (FNO)-based models, discussed next,  relieve the modeler of several of these drawbacks. Specifically, they possess the universal approximation property for continuous operators \citep{kovachki2023neural}; they can be easily adapted to different regular griddings of $\mathcal{D}$; and, they are extremely fast to fit and forecast with, by virtue of the Fast Fourier Transform.

\subsection{FNO-DST models for efficient computation}\label{subsec:FNO}

{Motivated by the discussion in the previous section about hidden dynamics, spatio-temporal models for which the underlying dynamics are not prescribed should, in general, include classes that are Markov of order greater than 1. Here we consider a flexible operator that maps sample paths of the spatio-temporal process $Y(\cdot, \cdot)$ from $L^2(\mathcal{D} \times [t_{k-\tau}, t_k];\mathbb{R})$ to $L^2(\mathcal{D} \times \{t_{k+h}\} ;\mathbb{R})$ for any $k \in \{\tau+1,\dots,T-h\}$. Note that the functional space of the input allows us to incorporate history-dependence into the operator, here by mixing convolutions over time. Specifically, at the core of our FNO defined below, is a space-time convolution of the form 
\begin{equation}\label{eq:linear_neural_operator}
v(\bfs, t_{k+h}) = \int_{t_{k - \tau}}^{t_k} \int_{\mathcal{D}} g(\bfs- \bfu, t_{k+h} - r; \bfgamma) v(\bfu,{r}) \, \text{d} \bfu \, \text{d} {r},
\end{equation}
where  \( v \in L^2(\mathcal{D} \times \mathcal{T}; \mathbb{R}) \) and $g(\cdot,\cdot;\bfgamma)$ is a Green's function parameterized by $\bfgamma$. Note from \eqref{eq:linear_neural_operator} that the operator is a convolution and hence time-invariant. Therefore, the time interval of the integral in \eqref{eq:linear_neural_operator} can be chosen to be $[0, \tau\delta]$, by shifting the time coordinate of $v(\cdot,\cdot)$:
\begin{equation}\label{eq:linear_neural_operator-2}
v(\bfs, t_{k+h}) = \int_{0}^{\tau \delta} \int_{\mathcal{D}} g(\bfs- \bfu, (h+\tau)\delta - q; \bfgamma) v_0^{(k)}(\bfu, q) \, \text{d} \bfu \, \text{d} {q},
\end{equation}
where the time-shifted spatio-temporal process $v_0^{(k)}(\bfu, t)$ is defined as
\[
v_0^{(k)}(\bfu, t) \equiv v(\bfu, t + t_{k-\tau}).
\]
Then $v(\bfs,t_{k+h}) \equiv \mathcal{G}_0[v_0^{(k)}]$, where $\mathcal{G}_0[\, \cdot \,]$ is the $k$-invariant operator defined from \eqref{eq:linear_neural_operator-2}. Without loss of generality we can therefore express the time-varying operator in \eqref{eq:linear_neural_operator} as a time-invariant operator on the shifted version of our spatio-temporal process, which corresponds to an implicit assumption that the underlying physical laws are temporally-invariant.

Now define $\mathcal{G}_0[\, \cdot \,] \equiv \mathcal{G}_{\bftheta}[\, \cdot \,]$, where $\bftheta$ are parameters that parameterize the FNO. The FNO uses a general (nonlinear) operator model for $\mathcal{G}_{\bftheta}$, with universal approximation properties as described in \citet{kovachki2023neural}:
\begin{equation}\label{eq:neural_operator-cont}
 \mathcal{G}_{\bftheta}[Y^{(k)}_0] = \mathcal{Q} \circ \sigma_{\text{a}}(\mathcal{W}_{L} + \mathcal{K}_L) \circ \cdots \circ \sigma_{\text{a}}(\mathcal{W}_{1} + \mathcal{K}_1) \circ \mathcal{P}[Y_0^{(k)}]; \quad Y_0^{(k)} \in L^2(\mathcal{D} \times [0,\tau\delta];\mathbb{R}),
\end{equation}
where $Y_0^{(k)}(\cdot, t) \equiv Y(\cdot, t+t_{k-\tau})$ for $t \in [0, \tau \delta]$.
Equation~\eqref{eq:neural_operator-cont} consists of three main components, which we now describe in detail. 
\begin{itemize}
    \item \textbf{Lifting operator $\mathcal{P}$:} The operator $\mathcal{P}[Y_0^{(k)}]$ is a pointwise mapping of $Y^{(k)}_0$'s graph into the higher-dimensional space $\mathcal{D}_v \subseteq \mathbb{R}^{d_v}$, where  $ d_v > d+2 $. Specifically, for $\bfx \equiv (\bfs^\top,t)^\top \in \mathcal{D} \times [0,\tau\delta]$, let $\bfx_{Y}^{(k)} \equiv (\bfx^\top, Y_0^{(k)}(\bfs,t))^\top$, which is a $(d+2)$-dimensional column vector. In our implementation, 
    \begin{equation}\label{eq:lifting}
    \bfv_0^{(k)}(\bfx) \equiv \mathcal{P}[Y_0^{(k)}](\bfx) = \mbf{H}_\mathcal{P} \bfx_{Y}^{(k)} +\mbf{b}_\mathcal{P};\quad \bfx \in \mathcal{D} \times [0,\tau\delta],
    \end{equation}
where \( \mbf{H}_\mathcal{P} \) is a \( d_v \times (d+2) \) matrix and \( \mbf{b}_\mathcal{P} \) is a $d_v$-dimensional vector. We collect the parameters in the operator $\mathcal{P}[\,\cdot\,]$ into the vector $\bfphi_{\mathcal{P}} \equiv \left(\text{vec}(\mbf{H}_\mathcal{P})^\top, \mbf{b}_\mathcal{P}^\top \right)^\top$.
\item \textbf{Kernel integral operator $\mathcal{K}_l$ and local linear operator $\mathcal{W}_l$:} For layers $l = 1, \dots, L,$ the operators $\mathcal{K}_l[\,\cdot\,]$ and $\mathcal{W}_l[\,\cdot\,]$ are the core components of the neural operator. Specifically, {$\mathcal{W}_l[\,\cdot\,]$ is the following linear transformation of the $d_v$-dimensional function $\bfv_{l-1}^{(k)}(\cdot)$, 
    \begin{equation}\label{eq:local_linear}
       \mathcal{W}_l[\bfv_{l-1}^{(k)}](\bfx) = \bfA_{\mathcal{W},l} \, \bfv_{l-1}^{(k)}(\bfx) + \bfb_{\mathcal{W},l};\quad \bfx \in \mathcal{D} \times [0,\tau\delta],
    \end{equation}
where $\bfA_{\mathcal{W},l}$ is a $d_v \times d_v$ matrix and $\bfb_{\mathcal{W},l}$ is a $d_v$-dimensional vector. We collect the parameters of $\mathcal{W}_l[\, \cdot \,]$ into the vector $\bfphi_{\mathcal{W},l} \equiv \left(\text{vec}(\bfA_{\mathcal{W},l})^\top,\bfb_{\mathcal{W},l}^\top\right)^\top$. The integral operator $\mathcal{K}_l[\,\cdot\,]$ is a multivariate version of a special case of \eqref{eq:linear_neural_operator-2} given by
    \begin{equation}\label{eq:integral_operator}
        \mathcal{K}_l[\bfv_{l-1}^{(k)}](\bfx) = \int_{\mathcal{D}\times[0,\tau\delta]}  {\bfG}_l(\bfx - \mbf{x}'; \bfphi_{\mathcal{K},l}) \, \bfv_{l-1}^{(k)}(\mbf{x}') \,  \intd \mbf{x}';\quad \bfx \in \mathcal{D} \times [0,\tau\delta],
    \end{equation}
    \noindent where $\bfG_l(\cdot;\bfphi_{\mathcal{K},l})$ is a $d_v \times d_v$ matrix of kernels with parameters collected into the vector $\bfphi_{\mathcal{K},l}$. Finally, the output of layer $l$, denoted  $\bfv_{l}^{(k)}$, is the $d_v$-dimensional function,}
    \begin{equation}\label{eq:neural_operator-cont-2}
        \bfv_{l}^{(k)}(\bfx) \equiv \ \sigma_{\text{a}}\left(\mathcal{W}_l[\bfv_{l-1}^{(k)}](\bfx) + \mathcal{K}_l[\bfv_{l-1}^{(k)}](\bfx)\right);\quad l = 1,\dots,L,\quad \bfx \in \mathcal{D} \times [0,\tau\delta], \\
    \end{equation}
    where recall that $\bfv_0^{(k)}(\bfx) \equiv \mathcal{P}[Y_0^{(k)}](\bfx)$ for some fixed $k$, and the map $\sigma_{\text{a}}(\cdot)$ is a pointwise activation function that renders the operator nonlinear. In our case, we use a rectified linear unit (ReLU) for $\sigma_{\text{a}}(\cdot)$, which returns its input vector with the negative components set to zero and with the positive components left unchanged.
    \item \textbf{Projection operator $\mathcal{Q}$:} This operator projects the $d_v$-dimensional output of the final layer back to the target output space. We implement this operator through
     \begin{equation}\label{eq:projection}
         \mathcal{Q}[\bfv_L^{(k)}](\bfx) =  \mbf{h}_{\mathcal{Q}}^\top \, \bfv_L^{(k)}(\bfx)   + b_{\mathcal{Q}};\quad \bfx \in \mathcal{D} \times [0,\tau\delta],
     \end{equation} 
     where \( \mbf{h}_{\mathcal{Q}} \) is a  \( d_v \)-dimensional vector, and \( b_{\mathcal{Q}}\) is a scalar. We collect the parameters in the operator $\mathcal{Q}[\,\cdot\,]$ into the vector $\bfphi_{\mathcal{Q}} \equiv (\mbf{h}_{\mathcal{Q}}^\top, b_{\mathcal{Q}})^\top$.
\end{itemize}

Recall that the neural operator given by \eqref{eq:neural_operator-cont} has universal approximation properties under suitable regulatory conditions. So too does the {Fourier} neural operator (FNO), where the operator $\mathcal{K}_l[\,\cdot\,]$ is expressed as a Fourier-inverse-Fourier pair. The Fourier transform of a convolution is a product of Fourier transforms, and hence,
\begin{align}\label{eq:Fourierpair}
\mathcal{K}_l\left[\bfv_{l-1}^{(k)}\right](\bfx) = \mathcal{F}^{-1}\big[\mathcal{F}[\bfG_l(\cdot; \bfphi_{\mathcal{K},l})] \mathcal{F}[\bfv_{l-1}^{(k)}]\big](\bfx),
\end{align}
where the Fourier operator and the inverse Fourier operator are
\begin{align}
  U(\bfomega) \equiv \mathcal{F}[u](\bfomega) &= \int_{\mathcal{D}\times[0,\tau\delta]}u(\bfx)\exp(-\iota \bfomega^\top \bfx) \intd \bfx;\quad \bfomega \in \mathbb{R}^{d+1}, \label{eq:Fourier1}\\
  \mathcal{F}^{-1}[U](\bfx) &= \frac{1}{(2\pi)^{d+1}}\int_{\mathbb{R}^{d+1}}U(\bfomega)\exp(\iota \bfomega^\top \bfx) \intd \bfomega; \quad \bfx \in {\mathcal{D}\times[0,\tau\delta]},\label{eq:Fourier2}
\end{align}
with $\bfomega$ denoting spatio-temporal angular frequency and $u(\cdot)$ a function in $L^2({\mathcal{D}\times[0,\tau\delta]};\mathbb{R})$. 
\paragraph{Definition:}\label{def:FNO}
Equation~\eqref{eq:neural_operator-cont}, together with the expression of 
$\mathcal{K}_l[\,\cdot\,]$ as a Fourier–inverse-Fourier transformation pair in \eqref{eq:Fourierpair}, 
defines the \emph{Fourier neural operator} (FNO) that maps functions in $L^2({\mathcal{D}\times[0,\tau\delta]};\mathbb{R})$ to functions in $L^2({\mathcal{D}\times \{(h+\tau)\delta\}};\mathbb{R})$.

Fast computation is key in any neural operator, and in the FNO we obtain it by using the Cooley-Tukey Fast Fourier Transform (FFT) \citep{tukey_fft}. Specifically, instead of using \eqref{eq:Fourier1} and \eqref{eq:Fourier2} we start with the discrete Fourier transform (DFT):
\begin{align}\label{eq:Fourier-disc1}
  U(\bfkappa) &= \sum_{\bfx \in \mathcal{D}^G \times\mathcal{T}_{\tau}^G} u(\bfx) \exp\left(-2\pi \iota \sum_{j=1}^{d+1}\frac{\kappa_j x_j}{m_j}\right);\quad \bfkappa \in \Omega,\\
  \label{eq:Fourier-disc2}
  u(\bfx) &= \frac{1}{m_1\cdots m_{d+1}} \sum_{\bfkappa \in \Omega} U(\bfkappa) \exp\left(2\pi \iota \sum_{j=1}^{d+1}\frac{\kappa_j x_j}{m_j}\right); \quad \bfx \in \mathcal{D}^G \times \mathcal{T}_{\tau}^G,
\end{align}
where $\mathcal{T}_{\tau}^G \equiv \{0, \delta, \dots, \tau \delta\}$; $m_j$ is the number of grid spacings in the the $j$-th spatio-temporal coordinate, so $m_{d+1} = \tau$; and  \(
\Omega \equiv \bigtimes_{j=1}^{d+1} ( 0, \dots, m_j-1)\).
Then, for computational efficiency, we shrink $\Omega$ to $\tilde{\Omega} \equiv \bigtimes_{j=1}^{d+1} ( 0, \dots, \tilde{m}_j-1)$, where $\tilde{m}_j \ll m_j, \ j = 1,\ldots d+1$. Hence, the product of Fourier transforms reduces to $\mathcal{O}(\prod_{j=1}^{d+1}\tilde{m}_j)$  matrix-vector multiplications of dimension $d_v$ that can be computed very quickly; see the Supplementary Material, Section S.1.3, where the FFT algorithm is briefly described.

In our implementation of the FNO, we do not declare a parametric form for \(\bfG_l(\,\cdot; {\bms{\phi}_{\mathcal{K},l}})\) in Equation \eqref{eq:Fourierpair} before finding its DFT. Instead, we define $\mathcal{F}[\bfG_l]$ in \eqref{eq:Fourierpair} as a spectral-weights tensor of complex numbers that is conjugately symmetric (since signals are taken to be real-valued), directly in the Fourier domain. We denote them by 
$\mathbf{F}_{\mathcal{K},l} \in \mathbb{C}^{d_v \times d_v \times \tilde{m}_1 \times \cdots \times \tilde{m}_{d+1}}$, 
in the high-dimensional space of complex numbers, which are trainable parameters that we collect into the vector
$\boldsymbol{\phi}_{\mathcal{K},l} \equiv \mathrm{vec}(\mathbf{F}_{\mathcal{K},l})$. They give the FNO the flexibility it needs to represent complex spatio-temporal dynamics. 

Let \( \boldsymbol{\theta} = \big( {\bfphi}_\mathcal{P}^\top,\,
\bfphi_{\mathcal{W},1}^\top, \, \bfphi_{\mathcal{K},1}^\top, \dots,\, \bfphi_{\mathcal{W},L}^\top,\, \bfphi_{\mathcal{K},L}^\top, \, \bfphi_{\mathcal{Q}}^\top \big)^\top \) be the set of trainable parameters associated with the neural operator in \eqref{eq:neural_operator-cont}.  Let ${\mbf{G}}(\,\cdot \, ; \,\bftheta)$ denote the neural operator implemented with the DFT \eqref{eq:Fourier-disc1}, which conveniently also admits a finite-dimensional parameterization. Now, for $k = \tau+1 , \dots, T-h,$ define the vector-valued stochastic time series $\bfY_{(k-\tau):k} \equiv \{\bfY_{k-\tau}, \dots, \bfY_{k}\}$ where the $|\mathcal{D}^G|$-dimensional vector $\bfY_{k}$ is $Y_k(\cdot)$ evaluated at $\mathcal{D}^G$. Then we can rewrite the output of the operator defined in \eqref{eq:neural_operator-cont} on the discretized domain as $\mbf{G}(\bfY_{(k-\tau):k};\,\bftheta)$. More details on these discretized tensors are discussed in the Supplementary Material, Section S.1.2. The discretized evaluation, \( \mbf{G}(\,\cdot\,; \, \bftheta) \), of the operator \(\mathcal{G}_{\bftheta}[\, \cdot \,]\) in \eqref{eq:neural_operator-cont} is a $|\mathcal{D}^G|$-dimensional vector that captures the spatio-temporal dynamics of a potentially complex physical process as follows: 
\begin{equation}
      \bfY_{k+h} = \mbf{G}(\bfY_{(k-\tau):k};\,\bftheta) + \bfeta_{k+h};\quad k = \tau+1,\,\dots, T-h.
    \label{eq:process_model_disc}
  \end{equation}
This should be compared to Equation \eqref{eq:IDE3}, where the operator $\bfG_{\bftheta''}(\bfY_k, \dots, \bfY_{k-\tau})\bfY_k$ is relatively constrained. Since the process model defined in \eqref{eq:process_model_disc} is a dynamic spatio-temporal (DST) model, we refer to this as the \textit{FNO-DST} model. 

We now turn to the additive terms $\{\bfeta_{\tau+1+h}, \dots, \bfeta_{T}\}$. We model the conditional covariance matrix $\bfSigma_{k+h}(\bfY_k; \bms{\alpha})$ of $\bfeta_{k+h}$ using a squared exponential covariance function with spatially and temporally varying standard deviation, given by
\begin{equation}
    \operatorname{cov}\left(\eta(\bfs,t_{k+h}),\eta(\bfs',t_{k+h}) \mid\bfY_k\right) = \sigma_{k+h}(\bfs;\bfY_k) \, \sigma_{k+h}(\bfs';\bfY_k) \, \exp\left( -\dfrac{\|\bfs - \bfs'\|^2}{2\alpha_{\text{r}}^2} \right),
    \label{eq:cov_function}
\end{equation}
for $\bfs, \bfs' \in \mathcal{D}^G.$ We cater for heteroscedasticity by modeling the standard deviation parameter at time $t_{k+h}$ as a function of the process $\bfY_k$. Specifically, we let
\begin{equation}
    \big(\sigma_{k+h}(\bfs_i;\bfY_k) : \bfs_i \in \mathcal{D}^G\big)^\top = f_{\text{NN}}(\bfY_{k}; \bms{\alpha}_{\text{NN}}),
\end{equation}
where $f_{\text{NN}}(\cdot)$ is a shallow feed-forward neural network that takes $\bfY_{k}$ as inputs and produces the vector of standard deviations across the spatial grid. This specifies $\bfSigma_{k+h}(\bfY_k; \bms{\alpha})$, where $\bms{\alpha} \equiv (\text{vec}(\bms{\alpha}_{\text{NN}})^\top, \alpha_{\text{r}})^\top$ are further trainable parameters in the FNO-DST model.

\subsection{Likelihood computation, training, and forecasting}\label{subsec:Inference}

{ Consider $N$ independent realisations from a spatio-temporal process model on $\mathcal{D} \times [0,K]$ generated using, for  example, random initial conditions or dynamical parameters. Assume that, for $n = 1,\dots, N$, we observe each process at time points $\{t_1, \dots, t_T\}$ with temporal step size $\delta$, to yield the discretized time series $\{\mathbf{Y}_1^{(n)}, \dots, \mathbf{Y}_T^{(n)}\}$.  Under the assumption of conditional independence given the history of $\tau$ observations, the likelihood function for the unknown parameters $(\boldsymbol{\theta}^\top, \boldsymbol{\alpha}^\top)^\top$ can be written as
\begin{equation}
\mathcal{L}(\boldsymbol{\theta}, \boldsymbol{\alpha}) = \prod_{n = 1}^N \prod_{k=\tau+1}^{T-h} p\big(\mathbf{Y}_{k+h}^{(n)} \mid \mathbf{Y}^{(n)}_{(k-\tau):k}, \boldsymbol{\theta}, \boldsymbol{\alpha} \big)
= \prod_{n = 1}^N \prod_{k=\tau+1}^{T-h}  \mathcal{L}_{k}^{(n)}(\boldsymbol{\theta}, \boldsymbol{\alpha}),
\label{eq:likelihood_function}
\end{equation}
where \(\mathbf{Y}^{(n)}_{(k-\tau) : k} \equiv \{ \bfY_{k-\tau}^{(n)},\, \dots,\, \bfY_{k}^{(n)} \}\). 
From the Gaussian VAR model given by \eqref{eq:process_model_disc},
\begin{equation}\label{eq:process_dist}
\resizebox{0.92\textwidth}{!}{$
    {\bfY}_{k+h}^{(n)} \mid \bfY^{(n)}_{(k-\tau):k}, \boldsymbol{\theta}, \boldsymbol{\alpha}
    \sim \text{Gau}\!\left(\mbf{G}(\bfY^{(n)}_{(k-\tau):k};\,\bftheta),
    {\bfSigma}_{k+h}^{(n)}(\bfY_k; \bms{\alpha})\right);
    n = 1,\dots, N, k =\tau+1, \dots, T-h,
$}
\end{equation}
where ${\bfSigma}_{k+h}^{(n)}(\bfY_k; \bms{\alpha})$ is the covariance matrix of $\bfeta_{k+h}^{(n)}$ parameterized by $\bms{\alpha}$. 
We maximize $\mathcal{L}(\boldsymbol{\theta}, \boldsymbol{\alpha})$ using a standard optimisation algorithm with an Adam learning schedule \citep{kingma2014adam}. 
Let $(\hat{\bftheta}, \hat{\bms{\alpha}})$ denote the estimated parameters that maximize $\mathcal{L}$. 
Substituting these estimates into Equation~\eqref{eq:process_dist} yields the forecasting distribution for replicate $i$:
\begin{equation}\label{eq:process_dist_estimated}
    \resizebox{0.92\textwidth}{!}{$
    {\bfY}_{k+h}^{(n)} \mid \bfY^{(n)}_{(k-\tau):k}, \hat{\bftheta}, \hat{\bms{\alpha}}
    \sim \text{Gau}(
        \mbf{G}(\bfY^{(n)}_{(k-\tau):k};\,\hat{\bftheta}),
        {\bfSigma}_{k+h}^{(n)}(\bfY_k; \hat{\bms{\alpha}})); n = 1,\dots, N, k = \tau+1,\dots, T-h,
        $}
\end{equation}
where $\mbf{G}(\bfY^{(n)}_{(k-\tau):k};\,\hat{\bftheta})$ is the estimated predictive mean and 
${\bfSigma}_{k+h}^{(n)}(\bfY_k; \hat{\bms{\alpha}})$ is the corresponding estimated predictive covariance matrix.

\section{Simulation experiments}\label{sec:experiments}

In this section, we conduct a comparative analysis of the forecasting performance of our proposed FNO-DST approach against current state-of-the-art approaches using two-dimensional (1-D space $\times$ time) simulations from Burgers' equation, a nonlinear PDE whose solution can be obtained numerically.  
The comparison uses performance metrics such as mean squared prediction error (MSPE), prediction interval coverage probability (PICP), and mean prediction interval width (MPIW); for more details on PICP and MPIW, see \cite{lakshminarayanan2017simple} and the Supplementary Material, Section S.2.2. In our simulation experiments, we compare forecasts from the FNO-DST model to those of the following: 
\begin{itemize}
    \item \textit{Convolutional Long Short-Term Memory} (ConvLSTM) neural network model \citep{NIPS2015_07563a3f}, which employs convolutional transitions within the inner LSTM module of the neural network. To train this model we replace $\mbf{G}(\, \cdot \,;\,\bftheta)$ in \eqref{eq:process_model_disc} with ConvLSTM blocks. 
    \item \textit{Space-Time DeepKriging} (STDK) model \citep{NAG2023100773}, which performs spatio-temporal prediction by learning the relationship between embedded spatio-temporal locations $(\bfs^\top,t)^\top,$ for $\bfs \in \mathcal{D}^G $ and $ t \in \{t_1, \dots, t_T\}$, and the underlying process $Y(\bfs, t)$, using neural networks. \cite{NAG2023100773} have shown that this approach outperforms another approach based on Gaussian regression with Vecchia’s approximation \citep{katzfuss2021general} for prediction of complex spatio-temporal processes.
    \item \textit{Persistence} model, which is a simple model where spatial observations from the last time point of the current sequence are used as forecasts for future time points at the corresponding locations. This approach can be thought of as providing a low-bar benchmark that other models' forecasts should improve upon.
    
\end{itemize}

Reproducible code is available from \url{https://github.com/pratiknag/FNO-DSTM-Code}.


 \subsection{Two-dimensional implementation}\label{subsec:simulation_description}

We consider the two-dimensional (1-D space $\times$ time) non-steady-state Burgers' equation \citep{hopf1950partial}, which is a nonlinear PDE that has been used in various applications such as modeling the one-dimensional flow of a viscous fluid. A solution to Burgers' equation can be obtained numerically, and can serve as a gold standard against which the forecasts from the four models given above can be compared. Burgers' equation for the evolution of a spatio-temporal field $\{u(s,t): s \in [0,1] , t \in [0,1]\}$ (i.e., $\mathcal{D} = [0,1]$ and $\mathcal{T} = [0,1]$), is written here as:
\begin{equation}
    \dfrac{\partial u(s,t)}{\partial t} + \dfrac{1}{2}\dfrac{\partial {u^2(s, t)} }{\partial s}= \gamma \dfrac{\partial^2 u(s,t)}{\partial s^2}; \quad (s,t) \in  (0, 1) \times (0, 1) ,
\label{eq:burgers_eqn}
\end{equation}
with initial condition $u(s, 0) = a_0(s), \, \text{for }s \in  [0, 1]$; and periodic boundary conditions $u(0,t) = u(1,t),$ for $t \in [0,1]$. Here \(\gamma \in [0,1]\) is the viscosity parameter. In order to solve \eqref{eq:burgers_eqn} for a given $\gamma$ and $a_0(\,\cdot\,)$, we rely on the Euler numerical method \citep{biswas2013discussion}, where the PDE is discretized onto a fine grid as follows: We replace temporal derivatives with a forward difference of step size $\Delta t = {1}/{10000}$, we replace spatial first-order derivatives with a backward difference of step size $\Delta s = {1}/{2^{11}}$, and  we replace spatial second-order derivatives with a symmetric second difference of the same step size. That is, the discretization of \eqref{eq:burgers_eqn} results in,
\begin{equation}
    \resizebox{0.22\hsize}{!}{$\frac{{u}(s_{i}, \tilde{t}_{j+1}) - {u}(s_{i}, \tilde{t}_{j})}{\Delta t}$} 
    + {u}(s_{i}, \tilde{t}_{j}) \resizebox{0.22\hsize}{!}{$\frac{{u}(s_{i}, \tilde{t}_{j}) - {u}(s_{i-1}, \tilde{t}_{j})}{\Delta s}$} = \gamma\resizebox{0.35\hsize}{!}{$ \frac{{u}(s_{i+1}, \tilde{t}_{j}) - 2{u}(s_{i}, \tilde{t}_{j}) + {u}(s_{i-1}, \tilde{t}_{j})}{(\Delta s)^2}$},
    \label{eq:discretization}
\end{equation}
with $s_0 = \tilde{t}_0 = 0$, periodic boundary conditions $u(s_0,\tilde{t}j) = u(s_{2^{11}},\tilde{t}_j)$, and a uniform grid $(s_i,\tilde{t}_j) = \left(i/2^{11},, j/10000\right)$ for $i = 1,\ldots,2^{11}$ and $j = 1,\ldots,10000$.
Hence Burgers' equation, discretized, can be rewritten as a one-step-ahead forecast by rearranging terms in \eqref{eq:discretization} as follows:
\begin{equation}
\begin{aligned}
    u(s_{i}, \tilde{t}_{j+1}) = u(s_{i}, \tilde{t}_{j}) - \frac{\Delta t}{\Delta s} u(s_{i}, \tilde{t}_{j}) 
    &\left( u(s_{i}, \tilde{t}_{j}) - u(s_{i-1}, \tilde{t}_{j}) \right) + \\
    &\gamma \frac{\Delta t}{(\Delta s)^2} 
    \left( u(s_{i+1}, \tilde{t}_{j}) - 2u(s_{i}, \tilde{t}_{j}) + u(s_{i-1}, \tilde{t}_{j}) \right),
\end{aligned}
    \label{eq:time_stepping}
\end{equation}
where the forecast at $\tilde{t}_{j+1}$ is made from local spatial knowledge at the current time $\tilde{t}_j$. (Note, we use $\tilde{t}_j$ for the simulator time point to distinguish it from the FNO-DST time point $t_j$.)
Since we have periodic boundary conditions, we simulate \( a_0(\cdot) \) on a unit circle once, from a wrapped Gaussian process with mean zero and covariance obtained from a chordal Mat\'ern covariance function \citep{GUINNESS2016143} with variance \( 1.0\), length scale \(1.0\), and smoothness \(2.0\). We consider two simulation settings: one has random $\gamma \equiv \gamma^{(1)} \sim \text{Uniform}(0.05, 0.7)$, and the other has fixed $\gamma \equiv \gamma^{(2)} = 0.4$. 

In this simulation study, we consider two different neural operators of the form
\begin{align*}
\mathcal{G}_{\bftheta} &: L^2([0,1]^2\times [0, \tau \delta]; \mathbb{R}) \mapsto L^2([0,1]^2\times \{(\tau + h) \delta\}; \mathbb{R}).
\end{align*}
We define the first neural operator, which we label FNO-DST-H (FNO-DST with `history'), to have $\tau = 4$. We define the second neural operator, which we label FNO-DST-NH (FNO-DST with `no history'), to have $\tau = 0$. For both operators we let $\delta = 0.1$, and $h = 5$ (5-step ahead forecast). We consider these two neural operators to give evidence to our intuition in Section \ref{subsec:IDE} that one needs to condition on past trajectories to sharpen the forecasts when the dynamics are variable (in the case of $\gamma^{(1)}$), even though \eqref{eq:discretization} suggests a Markov model. 

For each neural operator and simulation setting, we simulate $1000$  independent instances on the spatial domain $\mathcal{D}^G = \{ i/2^{11} \allowbreak : \allowbreak i=0,\dots,2^{11} \}$ with time step $\Delta_t = 1/10000$ using \eqref{eq:time_stepping}.  The neural operators use a coarser time step of $\delta = 0.1$.   Here, we let $\{t_1,\dots, t_{T}\} = \left\{ 0.1, 0.2, \dots, 1.0 \right\}$, and, we evaluate the forecast performance at $t_{10} = 1$ from information up to $t_5 = 0.5$. Note that $\mathcal{D}^{G}$ defines a uniform discretization of the unit interval with $2^{11}$ equispaced points, chosen to facilitate efficient FFT computations as discussed in the Supplementary Material, Section S.1.2.  Our temporal and spatial discretizations are used to construct the vectors $\{\bfY_k: k = 1,\dots,T\}$, which are needed for optimizing the log-likelihood defined in \eqref{eq:likelihood_function}. For each neural operator and simulation setting we use $N = 950$ instances for training; the remaining $N_0 = 50$ instances are used for testing.

\subsection{Simulation results}

\begin{figure}[htbp]
    \centering
    \vspace{-0.8em}
    \begin{subfigure}{0.87\textwidth}
        \centering
        \begin{tabular}{cc}
          FNO-DST-H & FNO-DST-NH \\
          \includegraphics[width=0.47\textwidth]{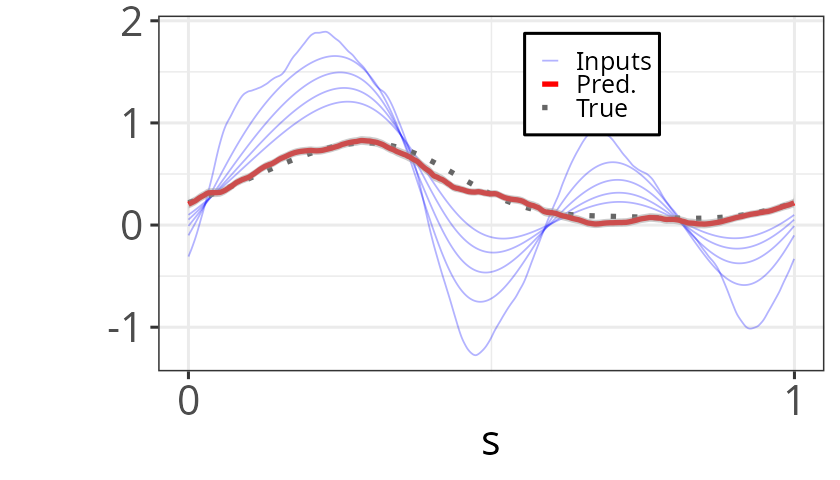} \vspace{2mm} & 
          \includegraphics[width=0.47\textwidth]{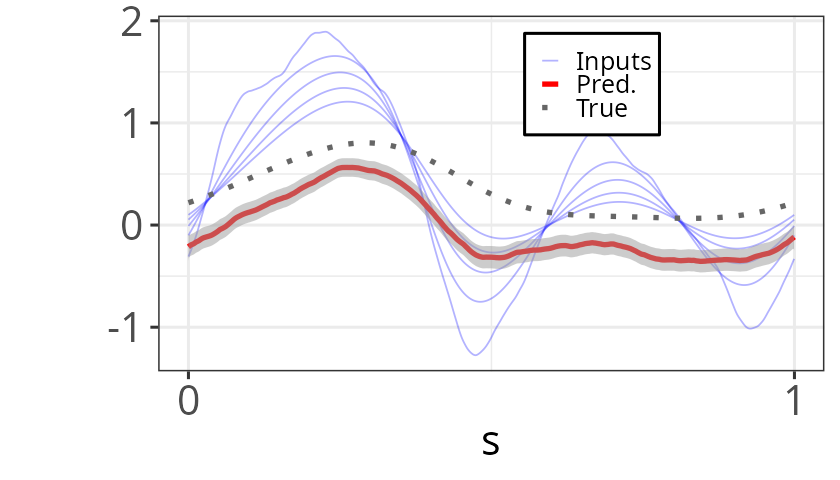} \vspace{2mm}  \\
          \includegraphics[width=0.45\textwidth]{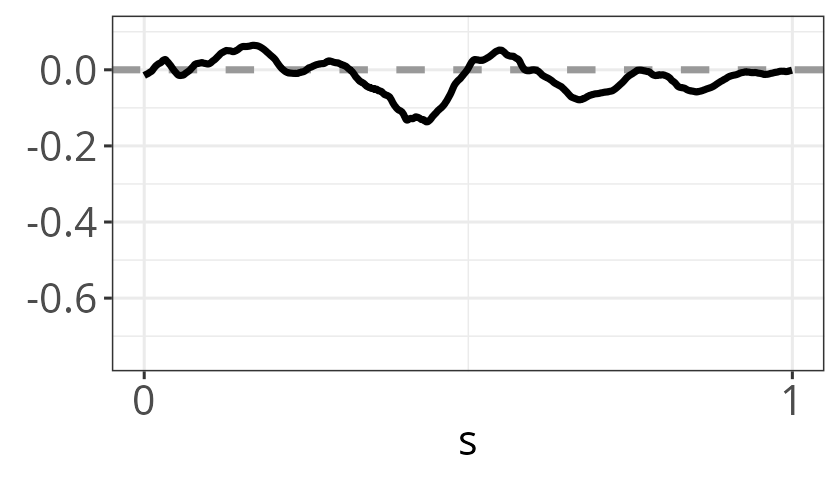} & 
          \includegraphics[width=0.45\textwidth]{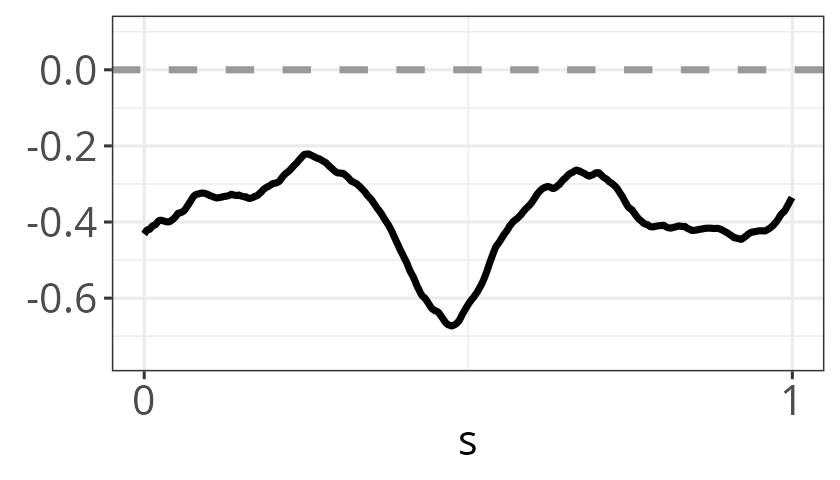} \\
        \end{tabular}
        \caption{Predicted fields (top row) and errors (bottom row) for random $\gamma = \gamma^{(1)}$.}
        \label{fig:burger-exp11}
    \end{subfigure}
    
    \vspace{-0.5em} 
    
    \begin{subfigure}{0.87\textwidth}
        \centering
        \begin{tabular}{cc}
          FNO-DST-H & FNO-DST-NH \\
          \includegraphics[width=0.47\textwidth]{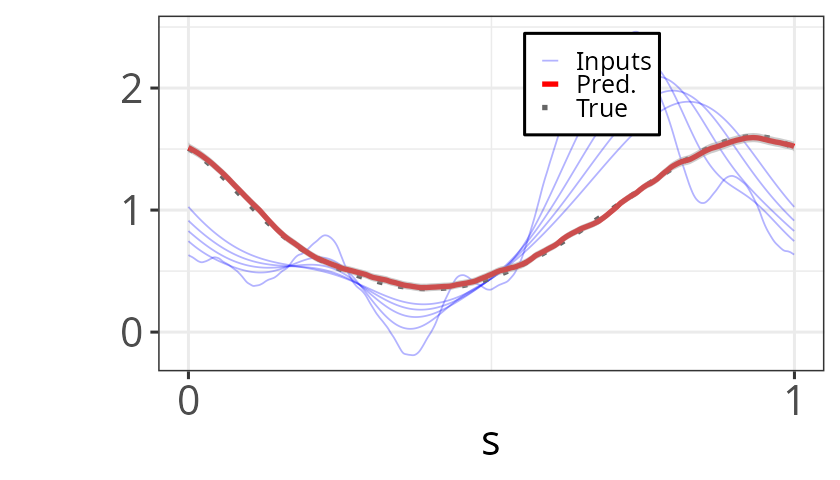} \vspace{2mm}& 
          \includegraphics[width=0.47\textwidth]{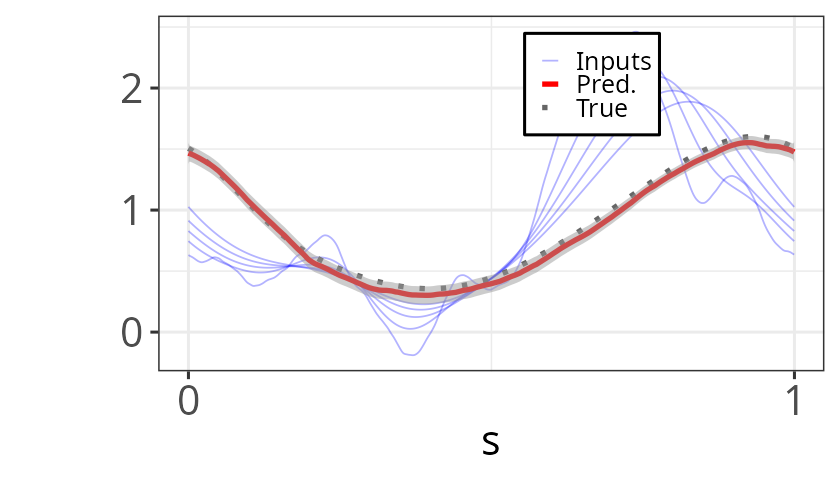} \vspace{2mm}\\
          \includegraphics[width=0.45\textwidth]{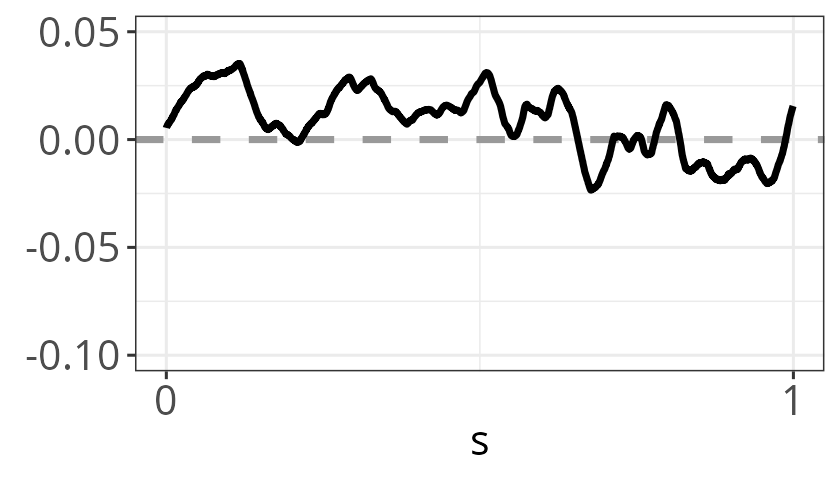} & 
          \includegraphics[width=0.45\textwidth]{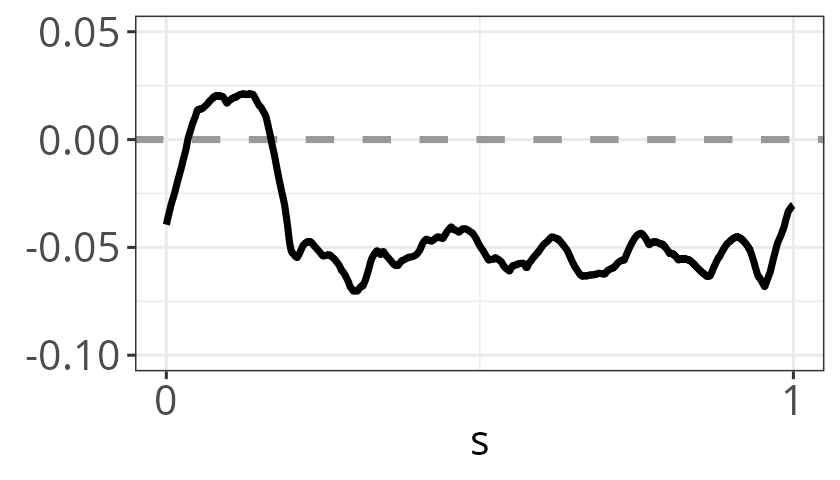} \\
        \end{tabular}
        \caption{Predicted fields (top row) and errors (bottom row) for fixed $\gamma = \gamma^{(2)}$.}
        \label{fig:burger-exp12}
    \end{subfigure}

    \caption{First testing instance out of $N_0=50$ testing instances for, respectively, forecasts from FNO-DST-H and FNO-DST-NH models at $T = 10$. (a): $\gamma = \gamma^{(1)}$, (b): $\gamma = \gamma^{(2)}$. In the top panels of (a) and (b), the current and past four steps (blue lines), the true spatial field at $T=10$ (black dotted line), and the forecasts (red line) with 95\% prediction intervals (gray shaded area), are shown. }
    \label{fig:burger-both}
\end{figure}

Figure~\ref{fig:burger-both} presents a comparison between FNO-DST-H and FNO-DST-NH for the first testing instance with random $\gamma = \gamma^{(1)}$ and the first testing instance with fixed $\gamma = \gamma^{(2)}$. In the Supplementary Material, Section S.2.2, the same types of comparisons are provided for two additional testing instances, which show patterns consistent with this first testing instance. 

Table~\ref{tab:burger-1d-2d} provides a numerical comparison between FNO-DST-H and FNO-DST-NH using forecasting performance metrics MSPE, PICP, and MPIW. These metrics are obtained by averaging over replications, space, and forecast times (See the Supplementary Material, Section S.2.1) at $t=1$.
From the comparative table and plots, it can be observed that  with $\gamma $ given by random $\gamma^{(1)}$, the inputs from the current and past $\tau=4$ steps during training enables FNO-DST-H to learn more effectively the underlying governing dynamics and produce a superior 5-step-ahead forecast of the spatial field $u(\cdot, 1)$ when compared to FNO-DST-NH. For $\gamma $ given by fixed $\gamma^{(2)}$, training on a single current value gives comparable results to training on the current and past four values. This supports our argument in Section \ref{subsec:IDE}, advocating for the inclusion of temporal history rather than relying on a single snapshot for forecasting a spatial field when the underlying PDE parameters are variable.

\begin{table}[t!]
    \centering
    \caption{Performance metrics for comparing FNO-DST-H  and FNO-DST-NH  for random $\gamma^{(1)}$ and fixed $\gamma^{(2)}$ in the simulation experiment of Section \ref{subsec:simulation_description}.}
    \begin{tabular}{|llccc|}
        \hline
       & Model & MSPE & PICP & MPIW \\
        \hline
        \hline
        \multirow{2}{*}{$\gamma ={\gamma^{(1)}}$} 
            & FNO-DST-NH     & 0.381 & 0.122  & 0.202 \\
            & FNO-DST-H  & 0.001 & 0.97    & 0.088 \\
        \hline
        \multirow{2}{*}{$\gamma = {\gamma^{(2)}}$} 
            & FNO-DST-NH    & 0.001     & 0.97    & 0.122     \\
            & FNO-DST-H  & 0.0007     & 0.98    & 0.065    \\
        \hline
    \end{tabular}
    \label{tab:burger-1d-2d}
\end{table}

To conclude this section, we compare the FNO-DST-H model with the three competing models given in the introduction to this section for simulation setting $\gamma = \gamma^{(1)}$. The comparison is shown visually in Figure~\ref{fig:burger} and using the forecasting performance metrics MSPE, PICP, and MPIW in Table \ref{tab:burger}. From the results, it can be seen that the FNO-DST-H produces valid forecasts (as do ConvLSTM and STDK), but it is more accurate, and it yields an MPIW that is smaller by an order of magnitude.

\begin{figure}[ht!]
    \centering
    \resizebox{\textwidth}{!}{
    \begin{tabular}{cccc}
        FNO-DST-H & ConvLSTM & STDK & Persistence \\

        \includegraphics[width=0.24\textwidth]{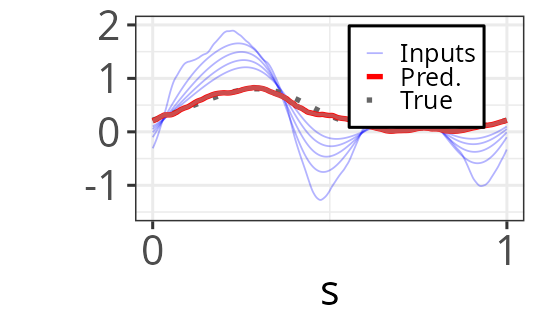} & 
        \includegraphics[width=0.24\textwidth]{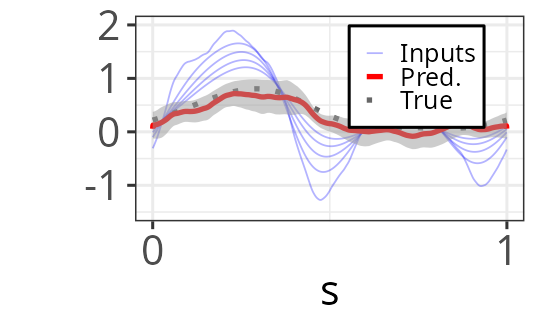} &
        \includegraphics[width=0.24\textwidth]{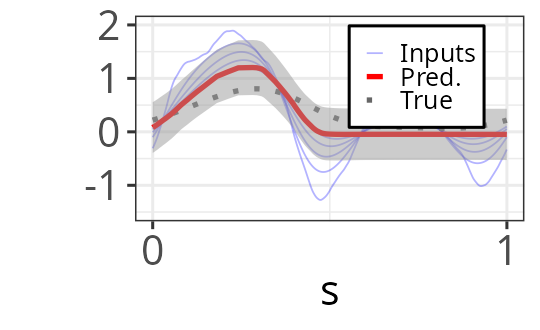} &
        \includegraphics[width=0.24\textwidth]{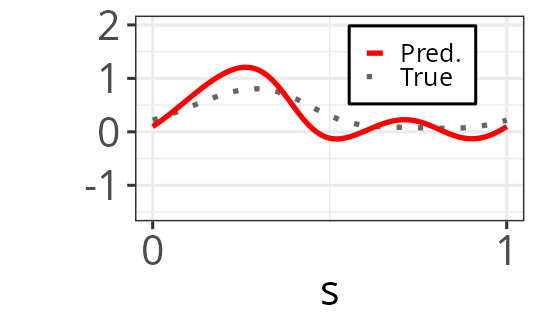} \\

        \includegraphics[width=0.23\textwidth]{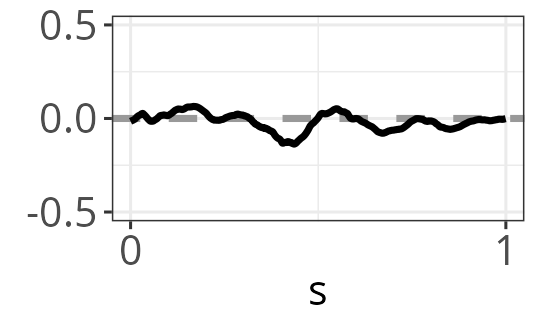} & 
        \includegraphics[width=0.23\textwidth]{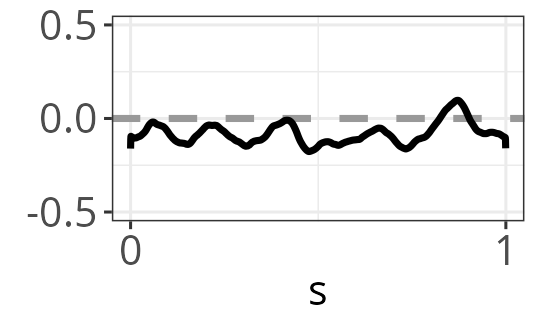} &
        \includegraphics[width=0.23\textwidth]{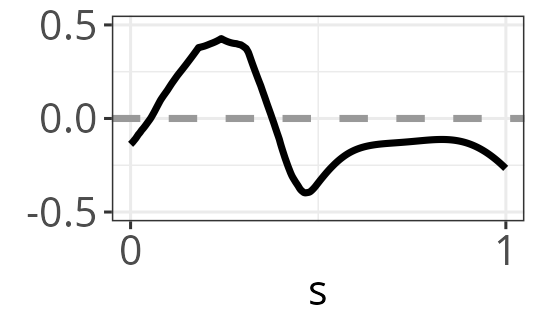} &
        \includegraphics[width=0.23\textwidth]{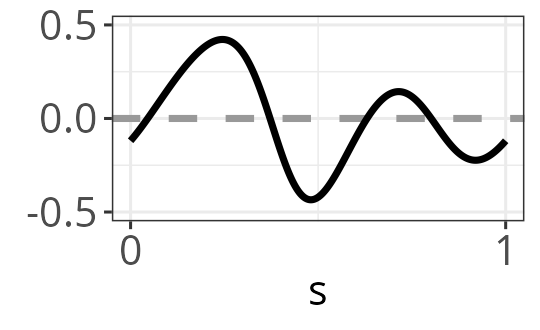} \\
    \end{tabular}
    }
    \caption{Forecast fields (top row by model) and corresponding prediction errors (bottom row by model) for the first testing instance out of $N_0=50$ testing instances. This visualization illustrates both the forecast quality at $T=10$ and error distribution across space for the simulation setting with random $\gamma = \gamma^{(1)}$. In the top panels, the current and past four steps (blue lines), the true spatial field at $T=10$ (black dotted line), and the forecasts (red line) with 95\% prediction intervals (gray shaded area), are shown.  }
    \label{fig:burger}
\end{figure}

\begin{table}[ht!]
\caption{Performance metrics for competing models for the simulation setting $\gamma = \gamma^{(1)}$ and FNO-DST-H, described in Section~\ref{subsec:simulation_description}.}
    \centering
    \begin{tabular}{|lccc|}
        \hline
        Model & MSPE & PICP & MPIW \\
        \hline
        \hline
        FNO-DST-H   & 0.001 & 0.97 & 0.088\\
        ConvLSTM & 0.054 & 0.98 & 0.574 \\
        STDK     & 0.075 & 0.99 &  1.013\\
        Persistence & 0.089 & -& -\\
        
        \hline
    \end{tabular}
    \label{tab:burger}
\end{table}

\section{FNO-DST forecasting in geophysical applications}\label{sec:application}
In this section, we conduct a comparative analysis of FNO-DST forecasting applied to two geophysical data sets. We also compare these forecasts with those obtained from another dynamical forecasting approach based on the physically motivated integro-difference equation discussed in Section \ref{subsec:IDE}. A neural-network version of the linear integro-difference equation \citep[CNN-IDE;][]{ZAMMITMANGION2020100408} is competitive when those dynamics are expected (as in SST forecasting), but it performs less well when little is known about the underlying dynamics.

\subsection{Application to SST data}\label{subsec:sst}

In this study, we use sea surface temperature (SST) data generated by the NEMO (Nucleus for European Modeling of the Ocean) simulation framework \citep{madec2024nemo}.
NEMO is an advanced modeling system designed for oceanic simulations, employing physically motivated equations to model both regional and global ocean circulation. As in climate reanalysis, it also employs historical observations to produce an assimilated representation of oceanic states, which ensures that its data closely align with actual SST measurements.  

We utilize NEMO SST data obtained from the \url{GLOBAL_ANALYSIS_FORECAST_PHY_001_024} product, which is provided by the Copernicus Marine Environment Monitoring Service (CMEMS). From this source, we aggregated the data onto a $0.5$-degree longitude-latitude grid. 
Following \cite{ZAMMITMANGION2020100408}, we concentrated our analysis and forecasting of SST on the North Atlantic Ocean. The region is divided into 19 distinct zones, each spanning a standardized $64 \times 64$ grid $\mathcal{D}^G \equiv \left\{ \left( \frac{i-1}{63}, \frac{j-1}{63} \right) : i,j \in \{1,\ldots,64\} \right\}
$. 

In this study, we focus on leveraging the FNO-DST framework for modeling the operator mapping defined in Section \ref{subsec:FNO} as
\begin{align*}
\mathcal{G}_{\bftheta} &: L^2([0,1]^2\times [0, \tau \delta]; \mathbb{R}) \mapsto L^2([0,1]^2\times \{(\tau + h) \delta\}; \mathbb{R}).
\end{align*}
To train our models, we use the initial $4000$ days of data available in this data product, covering the period from 27 December 2006 to 16 December 2017, with the 19 zones treated as replicates. The final 10 days of data are reserved for testing. Consequently there are $4000 \times 19 = 76000$ images that are used for training purposes. Figure \ref{fig:zone_partition} \citep[motivated from figures in][]{ZAMMITMANGION2020100408} 
\begin{figure}[ht!]
    \centering
    
     \includegraphics[width=\textwidth]{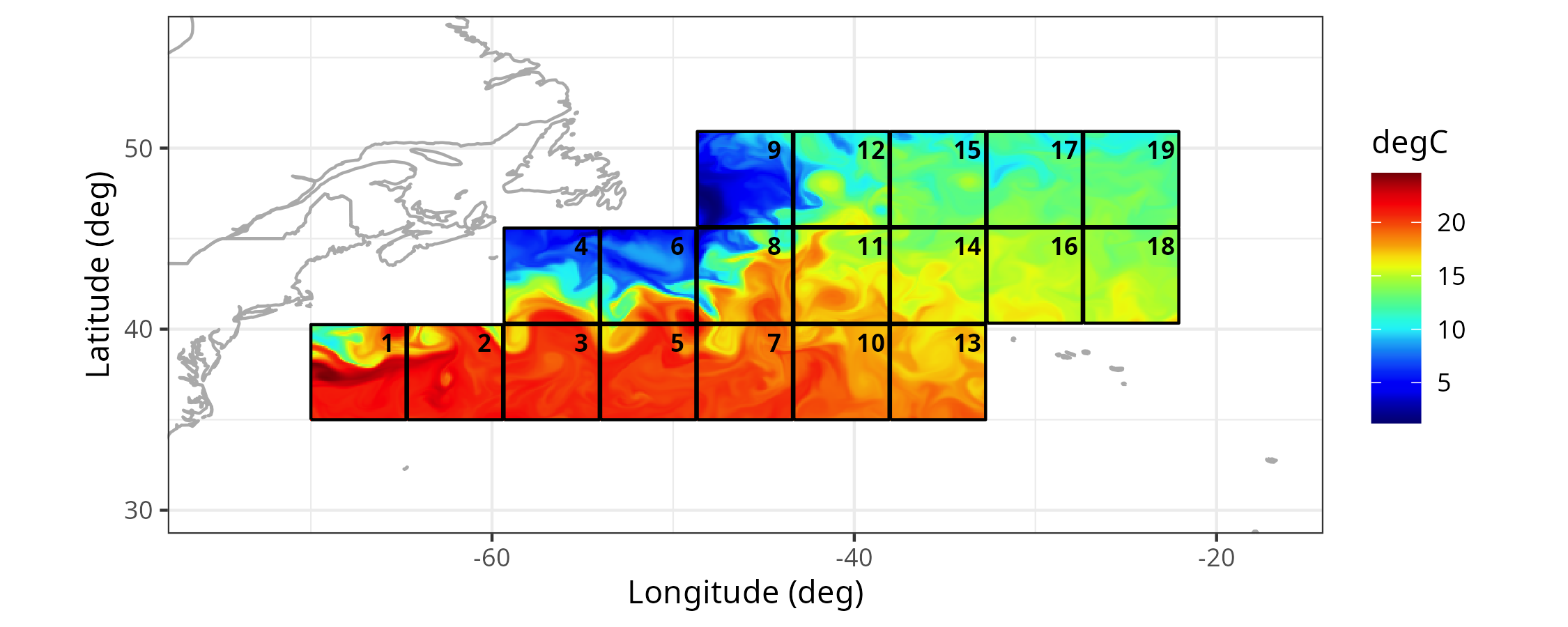}
    \caption{ North Atlantic Ocean sea surface temperature (SST) NEMO data on December 27, 2006, with 19 zones numbered and delineated by the black boxes.}
    \label{fig:zone_partition}
\end{figure}
presents the delineation of these 19 zones, illustrated with the SST data on December 27, 2006 (the first day in the data product). This subdivision into the 19 zones is primarily for computational efficiency; however, it does not significantly affect the approach, as the dominant dynamics within each zone can still be used to forecast the future from the recent past. We take $\tau = 2$ (recent past), $h = 3$ (3-step-ahead forecast) and discretize the temporal dimension to the daily resolution. Hence we forecast 3 days ahead from the current day and two preceding days.

Figure \ref{fig:sst} gives a visual comparison of the 3-day-ahead forecasts of the FNO-DST model and the CNN-IDE model. It is well established that SST data sets exhibit advection-diffusion properties, which the CNN-IDE model effectively incorporates through prior physical knowledge. Table \ref{tab:sst} uses the same metrics used in Tables \ref{tab:burger-1d-2d} and \ref{tab:burger} (Section \ref{sec:experiments}), showing an expanded comparison that includes the other three forecasts (ConvLSTM, STDK, and Persistence) that were compared in the simulations. While CNN-IDE forecasts of SST were best, the FNO-DST forecasts still performed well, even though no explicit geophysical knowledge was incorporated. This demonstrates its forecasting potential for applications where the dynamics are scarcely known, as is the case for the daily precipitation data in the next subsection.
\begin{figure}[t]
    \centering
    \resizebox{0.9\textwidth}{!}{
    \begin{tabular}{c ccc}  
                            &
        \includegraphics[]{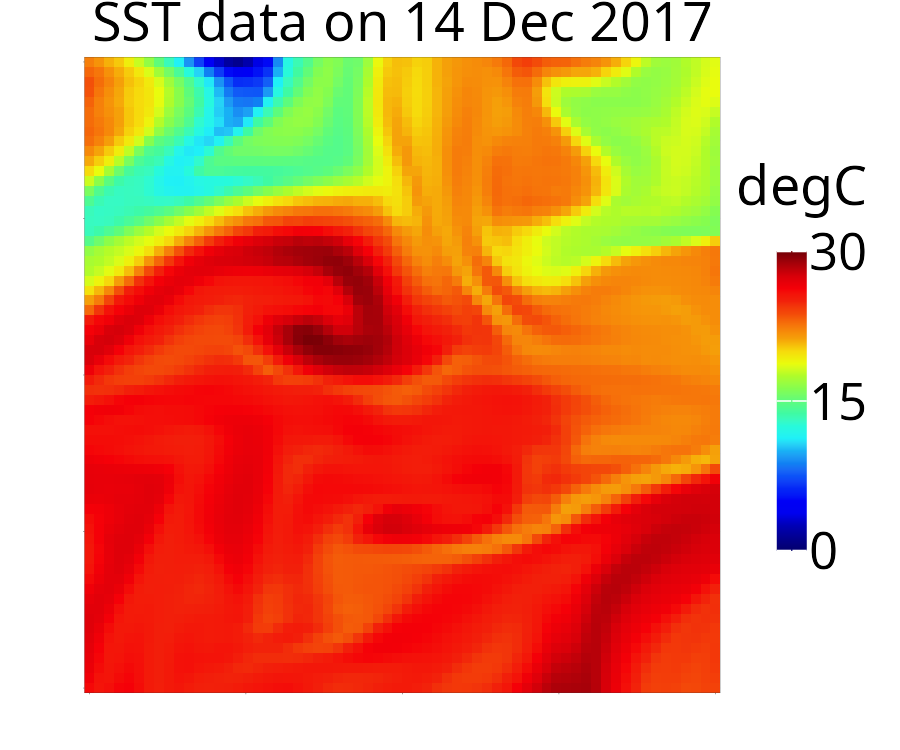} &
        \includegraphics[]{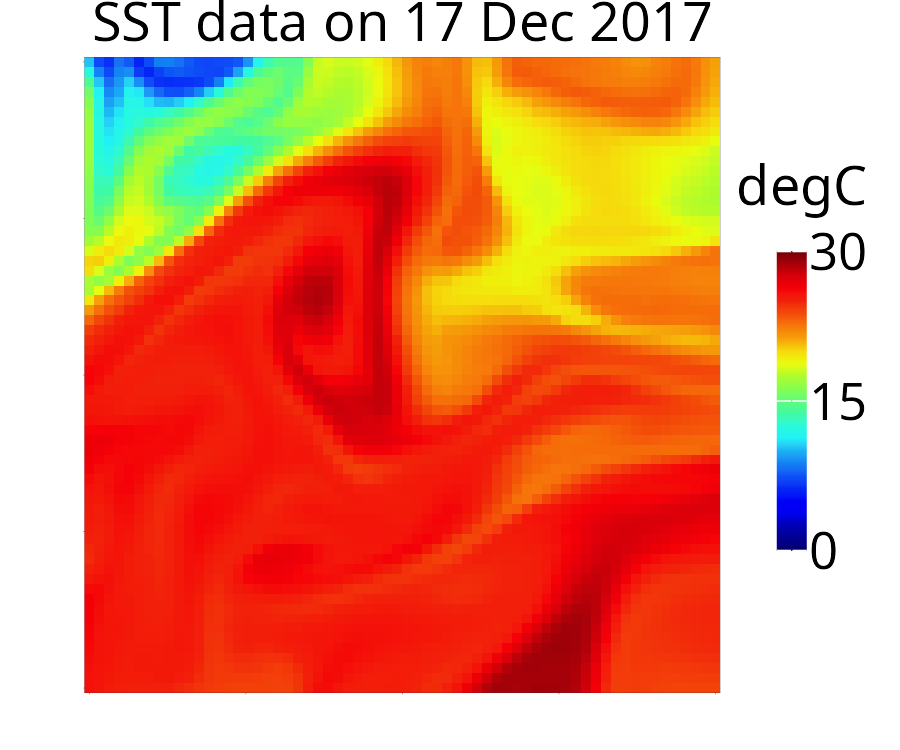} & \\

        \raisebox{10\height}{\centering \textbf{FNO-DST}} &
        \includegraphics[]{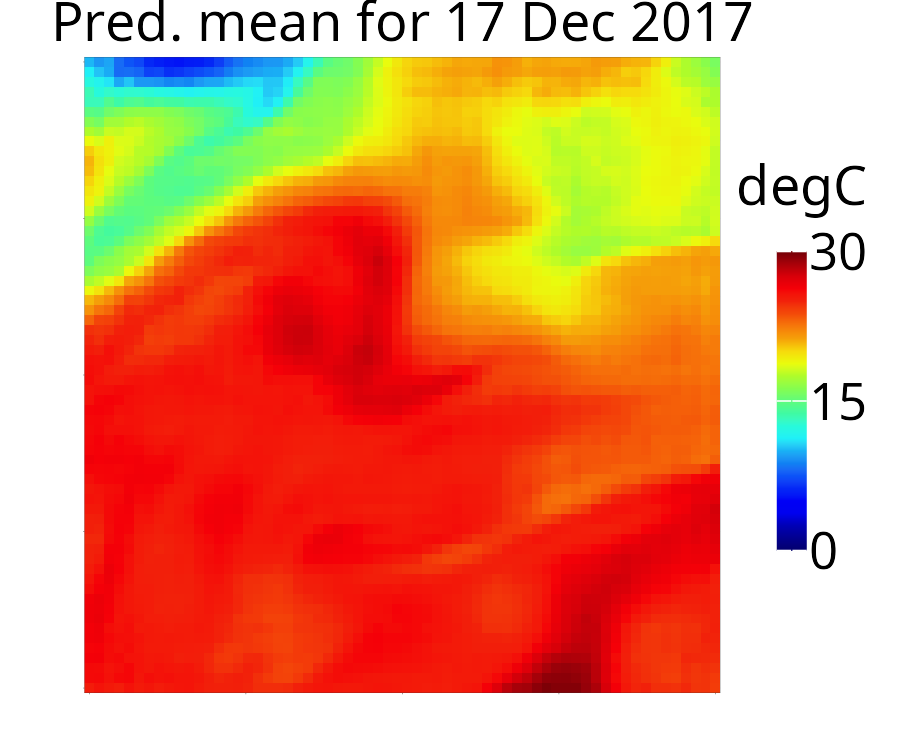} &
        \includegraphics[]{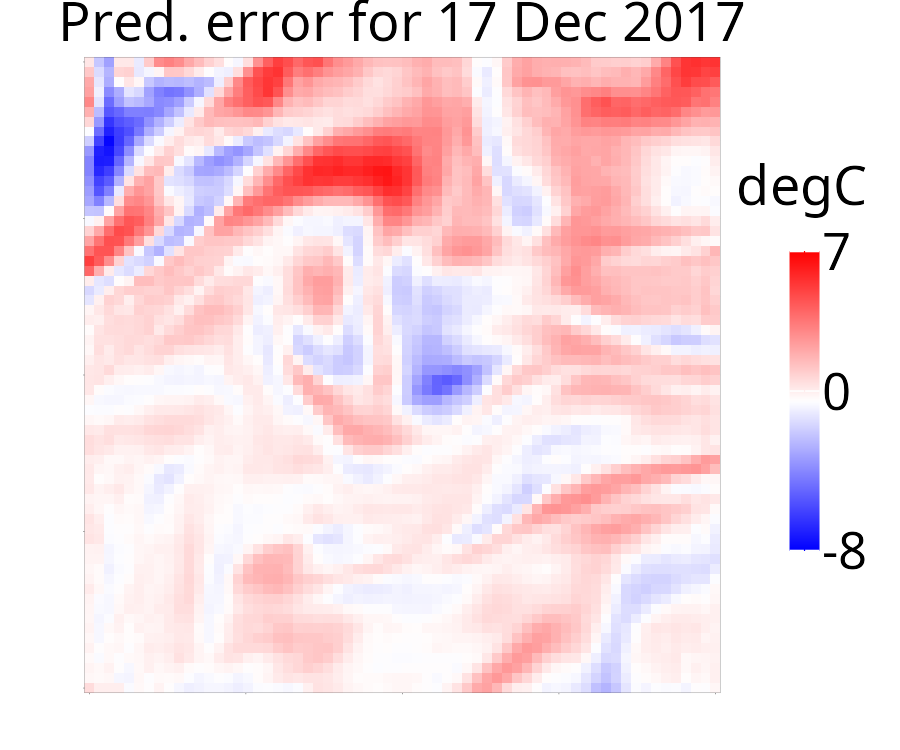} &
        \includegraphics[]{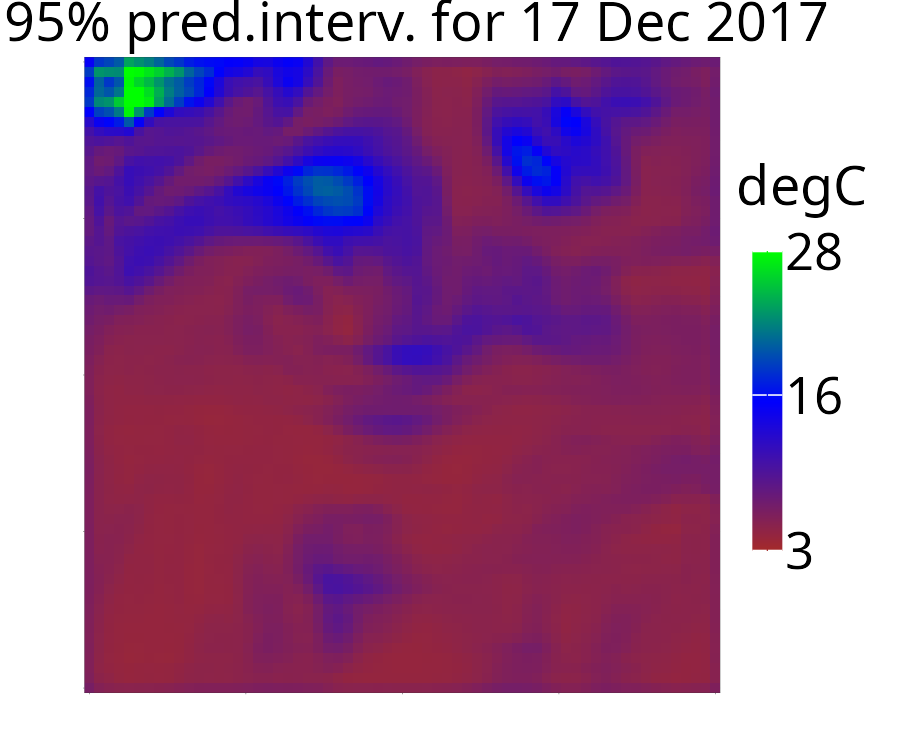} \\
        
        \raisebox{10\height}{\centering \textbf{CNN-IDE}} &
        \includegraphics[]{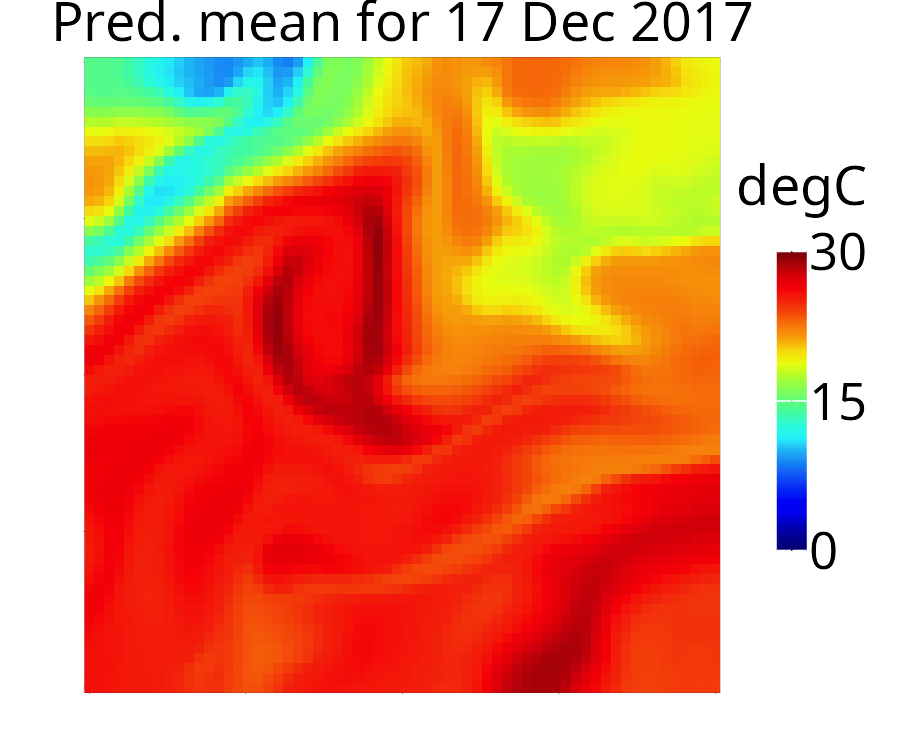} &
        \includegraphics[]{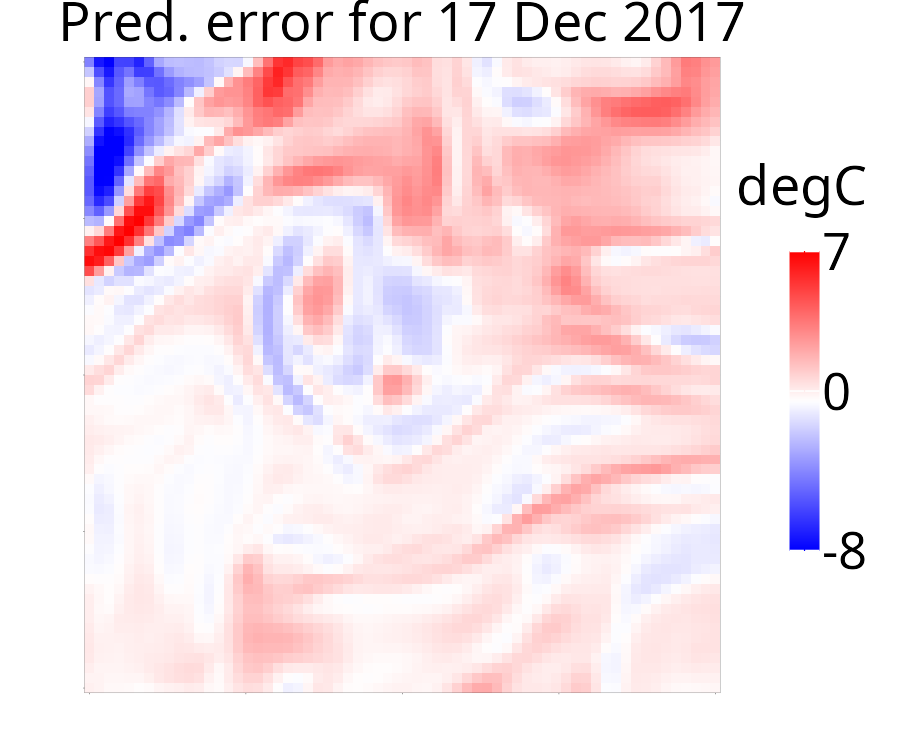} &
        \includegraphics[]{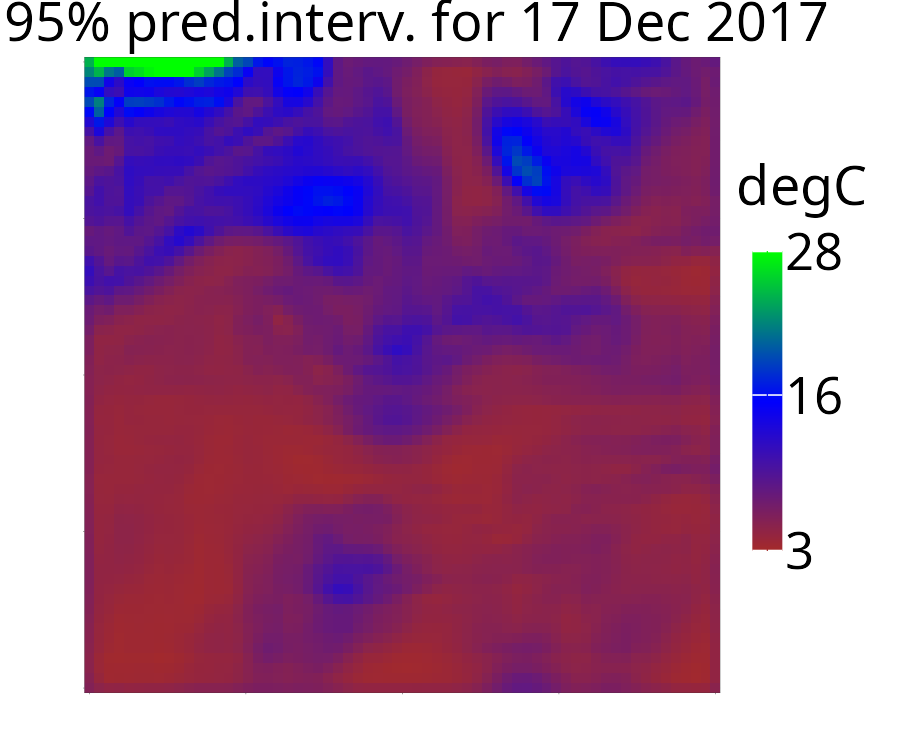} \\

     

    \end{tabular}
    }
    \caption{
        Visual comparison of FNO-DST and CNN-IDE forecasts for zone 3 in the North Atlantic Ocean. Forecasts of SST on 17 Dec 2017 were made from data on 12 Dec, 13 Dec, 14 Dec 2017. The top row shows 14 Dec 2017 data and its evolution to 17 Dec 2017, 3 days later. The middle row (FNO-DST) and bottom row (CNN-IDE) show the forecast results for 17 Dec 2017 (Pred. mean, Pred. error, and pixelwise 95\% Pred. interval width). 
    }
    \label{fig:sst}
\end{figure}

\begin{table}[h]
\caption{Forecasting performance metrics of SST forecasts for competing models described in Sections \ref{subsec:IDE} and \ref{sec:experiments}, averaged over 19 zones and over all 3-day-ahead forecasts from 17 December 2017 to 26 December 2017. (The FNO-DST model includes `history' from the current time back to the two previous times.) }
    \centering
    \begin{tabular}{|lccc|}
        \hline
        Model & MSPE & PICP & MPIW \\
        \hline
        \hline
        FNO-DST  & 1.56 & 0.95 & 5.77 \\
        CNN-IDE    & 1.34 & 0.95 & 5.65 \\
        ConvLSTM & 1.66 & 0.96 & 5.85 \\
        STDK     & 2.27 & 0.93 & 3.98 \\
        Persistence & 2.45 & - & - \\
        \hline
    \end{tabular}
    \label{tab:sst}
\end{table}

\newpage

\subsection{Application to precipitation data}\label{subsec:precip}

The second application uses daily precipitation data collected from weather stations distributed across Western and Central Europe \citep{toreti2014gridded}. The data set includes a comprehensive observational record of daily precipitation measurements spanning the period from 01 January 2014, to 31 December 2024. Figure~\ref{fig:precip-europe} illustrates the spatial distribution of the weather stations, where the subregion highlighted in red was  selected for our forecasting study.
To preprocess the data, a 10-day moving average was applied to a weather station's daily data to mitigate short-term fluctuations and fill in small gaps in the data. Each time series was then standardized to ensure consistent scaling across spatial locations and time points, with the sample mean and sample standard deviation saved in order to produce forecasts back on the original scale. Finally, spatio-temporal interpolation using the DeepKriging approach of \cite{NAG2023100773} was employed to generate precipitation fields on a $64\times 64$ standardized grid $\mathcal{D}^G \equiv \left\{ \left( \frac{i-1}{63}, \frac{j-1}{63} \right) : i,j \in \{1,\ldots,64\} \right\}$ defined over the highlighted subregion in Figure \ref{fig:precip-europe}, namely \(
[15^\circ\mathrm{E}, 25^\circ\mathrm{E}] \times [45^\circ\mathrm{N}, 55^\circ\mathrm{N}],
\) for all days between 01 January 2014 and 31 December 2024 inclusive. Further details on the dataset, its preprocessing, and the interpolation framework are provided in \cite{nag2025spatiotemporaldeepkrigingpytorchsupplementary}.
To train the models, we use the first 3200 days of the gridded spatial dataset, spanning the period from 1 January 2014 to 5 November 2023. The remaining 421 days, from 6 November 2023 to 31 December 2024, are reserved for testing. 
As in Section \ref{subsec:sst}, we take $\tau = 2$ (recent past), $h=3$ (3-step-ahead forecast). Note that, unlike the SST application that had 19 replicates, here there is only one replicate, and so training is on 3200 images.
\begin{figure}[t!]
    \centering
    \includegraphics[]{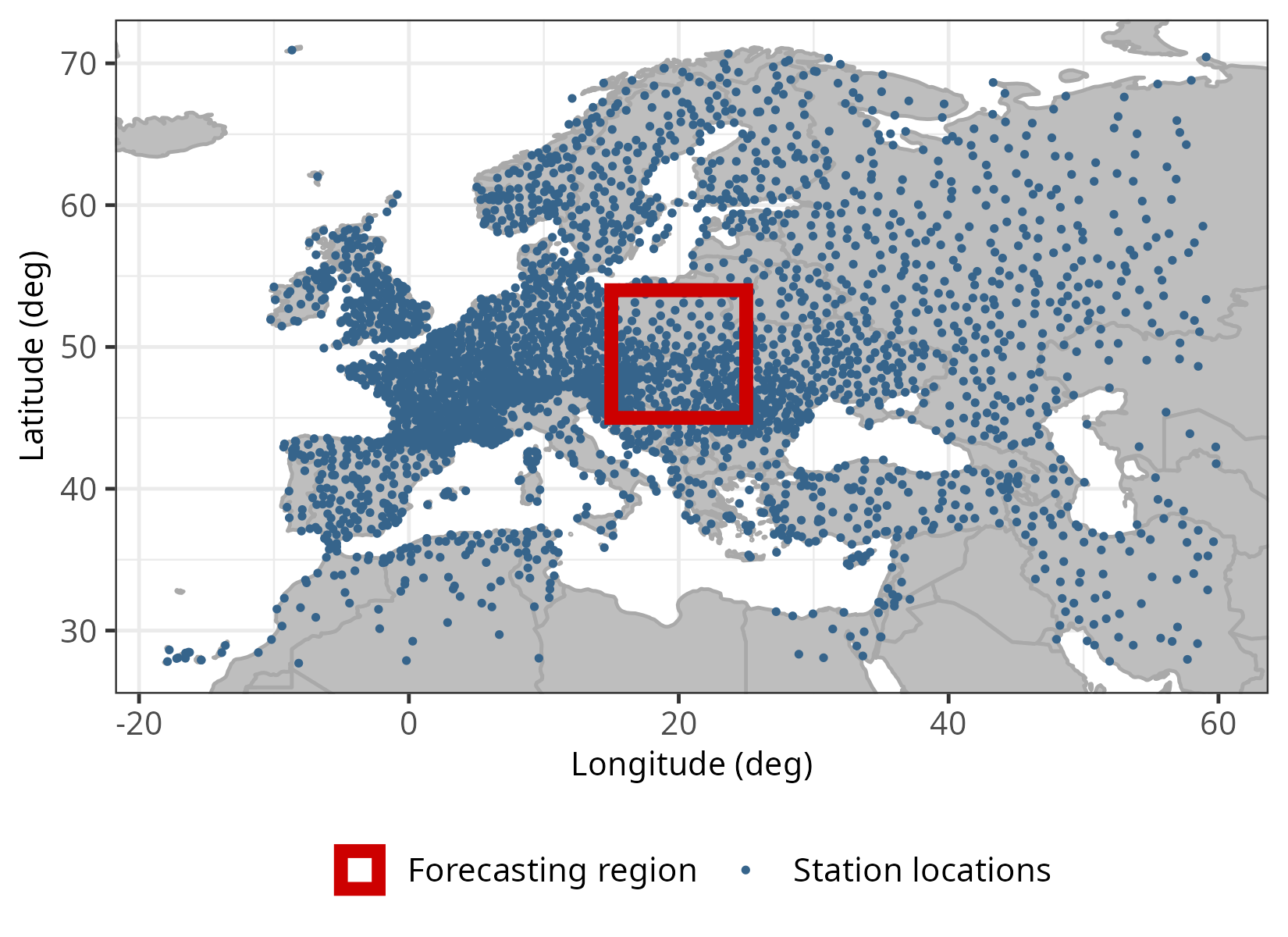}
    \caption{ Distribution of weather-station locations across western and central Europe. The region highlighted in red has been specifically chosen for forecasting purposes.}
    \label{fig:precip-europe}
\end{figure}

\begin{figure}[t]
    \centering
    \resizebox{0.9\textwidth}{!}{
    \begin{tabular}{c ccc}  
  
         &
        {\includegraphics[]{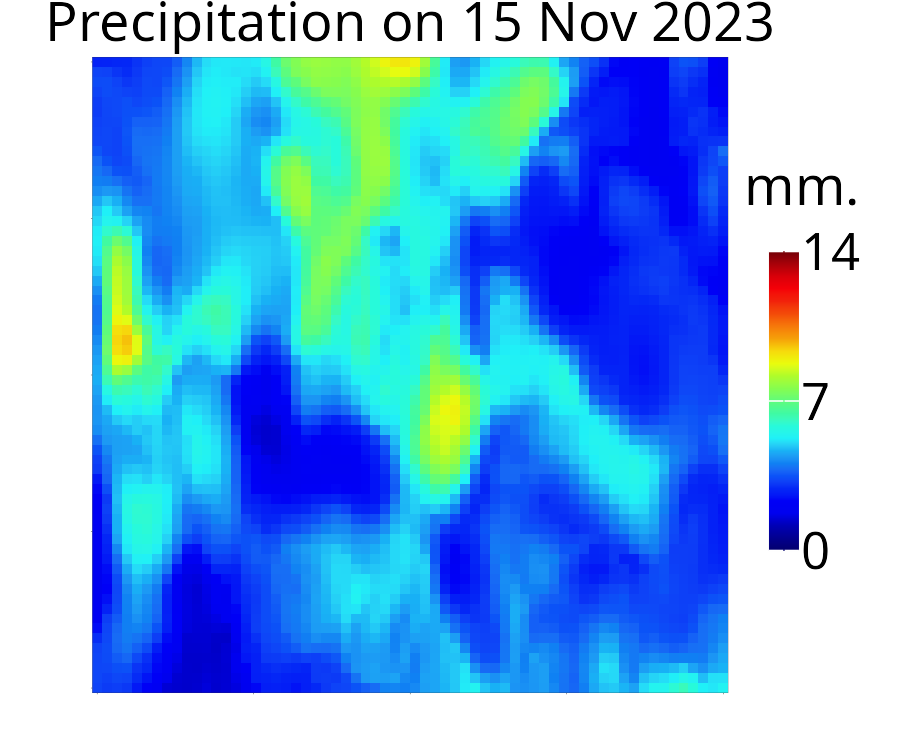}} &
       {\includegraphics[]{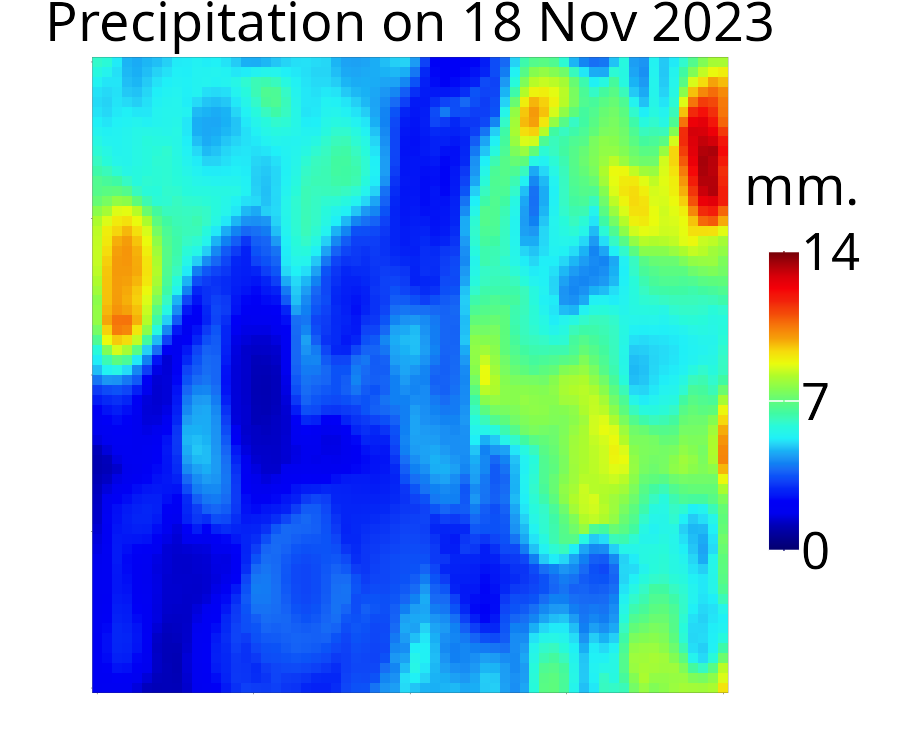}} & \\
    
        \raisebox{10\height}{\centering \textbf{FNO-DST}} & 
        {\includegraphics[]{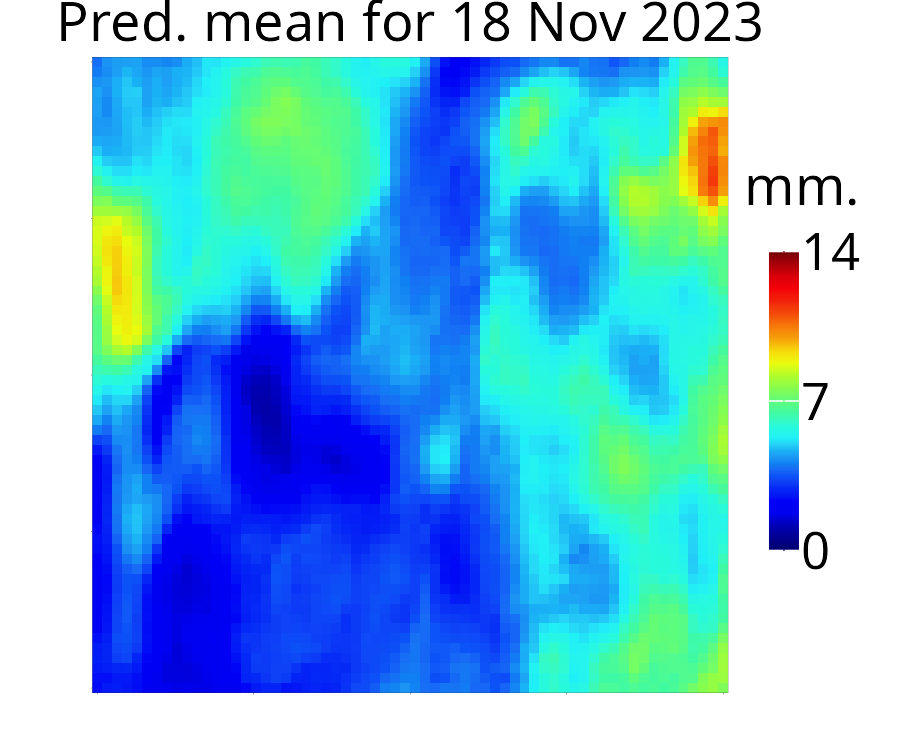}} &
       {\includegraphics[]{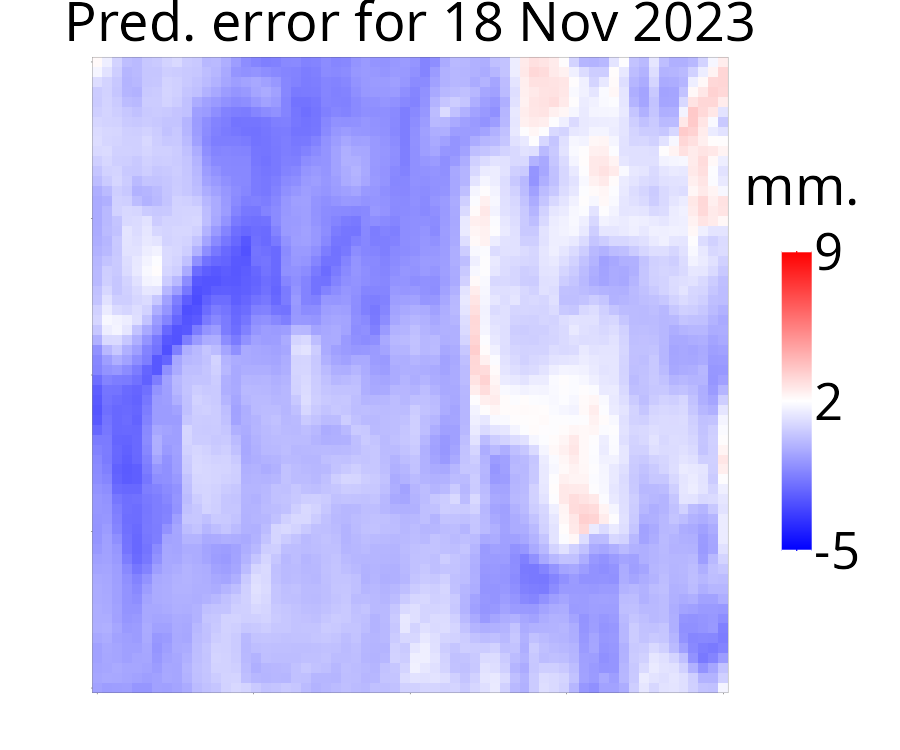}} &
      {\includegraphics[]{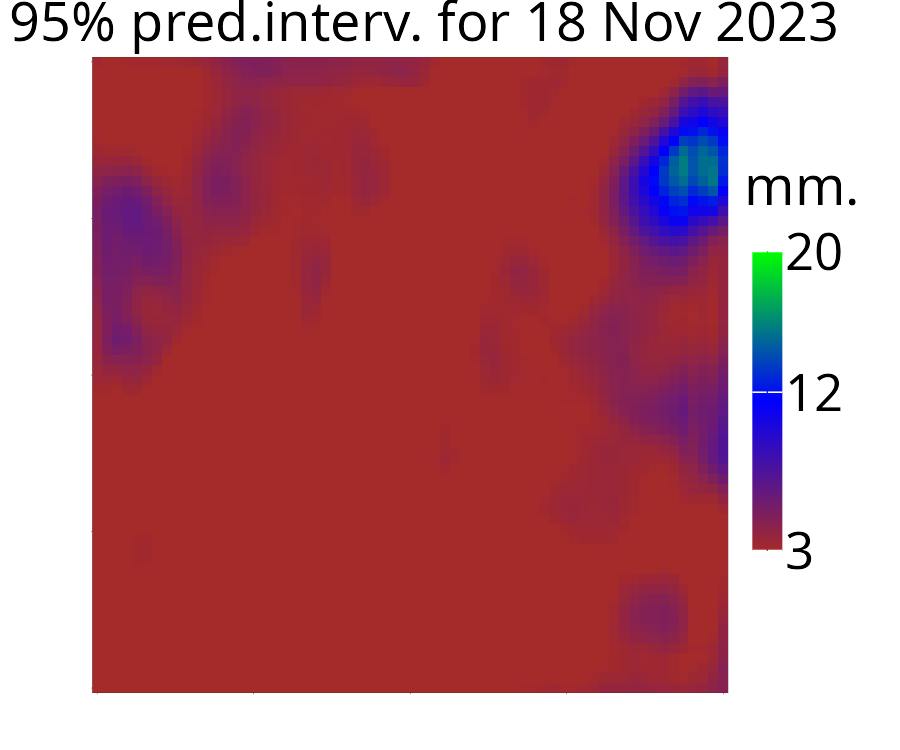}} \\  

      \raisebox{10\height}{\centering \textbf{CNN-IDE}} & 
        {\includegraphics[]{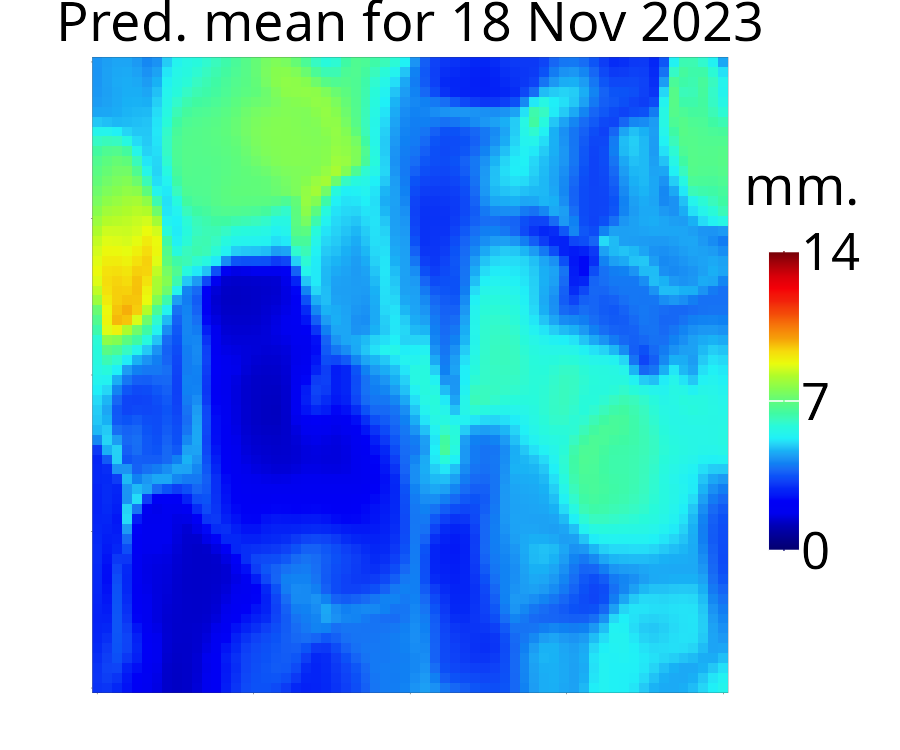}} &
       {\includegraphics[]{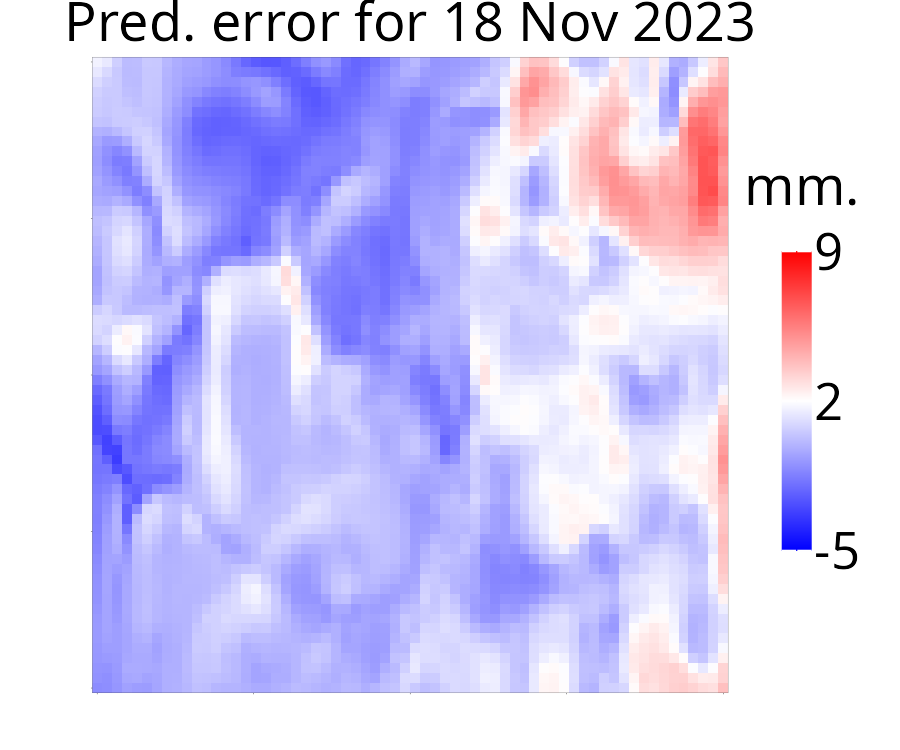}} &
        {\includegraphics[]{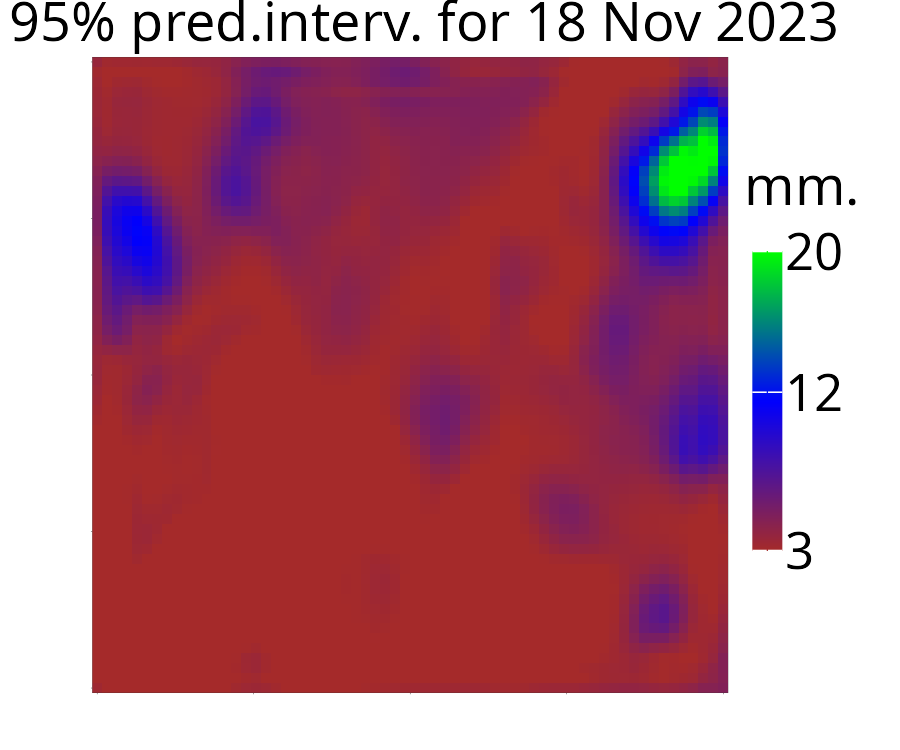}} \\
    \end{tabular}
    }
    \caption{Visual comparison of FNO-DST and CNN-IDE forecasts for the precipitation data over the highlighted region shown in Figure \ref{fig:precip-europe}. Forecasts on 18 Nov 2023 were made from data on 13 Nov, 14 Nov, 15 Nov 2023. The top row shows 15 Nov 2023 data and its evolution to 18 Nov 2023, 3 days later. The middle row (FNO-DST) and bottom row (CNN-IDE) show the forecast results (Pred. mean, Pred. error, and pixelwise 95\% Pred. interval width).}
    \label{fig:precip}
\end{figure}
\begin{table}[h]
\caption{Performance metrics of different competing models described in Sections \ref{sec:experiments} and \ref{sec:application}, averaged over all 3-step-ahead forecasts at all spatial locations and all days from 06 Nov 2023 to 31 Dec 2024. }
    \centering
    \begin{tabular}{|lccc|}
        \hline
        Model & MSPE & PICP & MPIW \\
        \hline
        \hline
        FNO-DST   & 0.32 & 0.96 & 1.95 \\
        CNN-IDE    & 0.46 & 0.96 & 2.43 \\
        ConvLSTM & 0.38 & 0.94 & 2.23 \\
        STDK     & 0.82 & 0.92 & 1.97 \\
        Persistence & 0.84 & - & - \\
        \hline
    \end{tabular}
    \label{tab:precip}
\end{table}
Figure~\ref{fig:precip} gives a visual comparison of the 3-day-ahead forecasts from the FNO-DST model and the CNN-IDE model. Table \ref{tab:precip} uses the same metrics and forecasting approaches as in Table \ref{tab:sst}. The results clearly demonstrate that the FNO-DST forecasting of precipitation outperforms the other models' forecasts across all evaluation metrics. For the precipitation application, the dynamics are relatively poorly understood, and CNN-IDE comes third in performance. As shown in Figure \ref{fig:precip}, the FNO-DST model is able to forecast an East–West precipitation front that is similar to what actually occurs on 18 November 2023. The CNN-IDE model, on the other hand, does not reproduce this feature. The simulation results in Section \ref{sec:experiments} and the applications in this section highlight the inferential and computational efficiency of FNO-DST forecasting in diverse settings.

\newpage 

\section{Conclusions}\label{sec:conclusion}

In this work, we develop a framework for dynamic spatio-temporal (DST) modeling where Fourier Neural Operators (FNOs) are used to forecast entire spatial fields in a computationally efficient manner. Discretized partial- and integro-differential-equation models often struggle to accommodate nonlinear interactions and complex dynamics inherent in many environmental and physical systems, and this affects their forecasting performance. Our approach leverages the FNO's ability to learn and then to approximate mappings between spatial fields across time using fast, parallelizable computations and without requiring explicit knowledge of the underlying dynamical mechanisms.

We evaluated the forecasting performance of the proposed FNO-DST model defined in Section \ref{subsec:FNO} through a combination of simulation studies and real-world applications. In the simulation experiments based on Burgers' equation, which is a nonlinear PDE, the FNO-DST model demonstrated strong forecasting skill in capturing the nonlinear dynamics governing the system. In the real-data applications, we applied the model to two distinct spatio-temporal geophysical settings: forecasting sea surface temperature (SST) over the North Atlantic Ocean, and forecasting precipitation over a region in Europe. In the SST-forecasting application, where physical PDE-based models have been previously developed, the FNO-DST model performed almost as well as a PDE-based model. In the precipitation-forecasting application, where the underlying dynamics are considerably more complex and less-well characterized, the FNO-DST model was able to learn the dynamics and provide superior forecasts according to several performance metrics. Our results demonstrate the ability of the FNO-based framework to adapt to a wide range of spatio-temporal systems for which the dynamics are unknown or little known.

Beyond its forecasting ability, a key strength of the FNO-based framework is its computational efficiency and scalability, both during training and post-training. This offers a substantial computational advantage for real-time forecasting applications, especially in high-dimensional domains, where traditional statistical and mechanistic models can become computationally prohibitive, or perhaps the traditional models are over-simplified for computational efficiency. Just as importantly, by embedding the FNO within a statistical model, the framework naturally accommodates uncertainty quantification, which is critical for decision making.

There remain several promising avenues for future research. First, extensions to incorporate physical constraints or conservation laws directly into the FNO framework will further improve forecasting skill while enhancing model interpretability. Second, developing a Bayesian hierarchical statistical formulation of the FNO-DST model will improve uncertainty quantification as data and parameter uncertainties are included. Finally, while the current work focuses on spatial lattices, adapting the methodology for irregular spatial domains, non-Euclidean spaces, or multivariate spatio-temporal processes represent important directions for broadening the applicability of the FNO-DST approach to forecasting.

\section*{Acknowledgement}
This material is based upon work supported by the Air Force Office of Scientific Research under award number FA2386-23-1-4100. Nag and Zammit-Mangion also acknowledge travel support from the JRC, Italy to visit their research centre, which motivated the application case study presented in Section \ref{subsec:precip}.


\section*{Appendix: Notation}

\begin{description}

\item[\(\mathcal{D}\):] Continuous spatial domain.
\item[\(\mathcal{T}\):] Continuous temporal domain.
\item[\(\mathbf{s}\):] The spatial coordinate, \(\mathbf{s} \in \mathcal{D}\).  
\item[\(t\):] The temporal coordinate, \(t \in \mathcal{T}\).  
\item[\(\mathcal{C}(\cdot, \cdot)\):] The deterministic real-valued continuous-time PDE-based spatial process.
\item[\(\bfgamma\):] Parameters associated with the underlying PDE.
\item[\(g(\cdot,\cdot),g_j(\cdot, \cdot)\):] Green's functions and kernels in different scenarios.
\item[\(\eta(\cdot,\cdot)\):] Spatially correlated random innovation term. 
\item[\(Y(\cdot,\cdot)\):] The spatio-temporal stochastic process being modeled.
\item[\(Y_k(\cdot) \equiv Y(\cdot, t_k)\):] The stochastic (spatial) process evaluated at time point $t_k$.
\item[\(\{Y_k(\cdot)\}_k \equiv \{Y(\bfs, t_k) : \bfs \in \mathcal{D}\}_{k=1}^{T}\):] The time series of spatial processes defined on the domain $\mathcal{D}$.
\item[\(\mathcal{D}^G\):] BAU-discretization of $\mathcal{D}$.
\item[\(\bfeta_{k+h}\):] $\eta(\cdot,t_{k+h})$ evaluated on $\mathcal{D}^G$.
\item[\(\bfSigma_{k+h}(\bfY_k; \bms{\alpha})\):] The covariance matrix associated with $\bfeta_{k+h}$, which is a function of $\bfY_k$ and parameterized by $\bms{\alpha}$.
\item[\(\bfY_k\):] $Y_k(\cdot)$ evaluated on $\mathcal{D}^G$.
\item[\(\bfG_{\bfgamma},\bfG_{j,\bftheta'},\bfG_{\bftheta''}\):] Matrices representing BAU-discretized Green's functions of the operators in 
\eqref{eq:IDE}, \eqref{eq:IDE_history}, and \eqref{eq:IDE3}, 
with parameters $\bfgamma$, $\bftheta'$, and $\bftheta''$, respectively.
\item[\(\tau\):] Past lag used to form the conditioning history. 
\item[\(h\):] Forecast horizon for h-step-ahead forecast.
\item[\(\mathbf{x} \equiv (\mathbf{s}^\top, t)^\top\):] The spatio-temporal coordinate, \(\mathbf{x} \in \mathcal{D} \times \mathcal{T}\).
\item[\(\mathcal{G}_{\bftheta}\):] The FNO operator parameterized by $\bftheta$.
\item[\(\mathcal{P}, \mathcal{K}, \mathcal{W}, \mathcal{Q}\):] 
The lifting operator, kernel integral operator, local linear operator, and projection operator, as defined in 
\eqref{eq:lifting}, \eqref{eq:integral_operator}, \eqref{eq:local_linear}, and \eqref{eq:projection}, respectively.
\item[\(\mathcal{F}, \mathcal{F}^{-1}\):] 
The Fourier and inverse Fourier operators, as defined in \eqref{eq:Fourier1} and \eqref{eq:Fourier2}, respectively.

\item[\(\mathbf{G}(\cdot;\bftheta)\):] 
Vector representing the BAU-discretized output of the operator \(\mathcal{G}_{\bftheta}\).

\item[\(\bfY_{(k-\tau):k}\):] Vector-valued stochastic time series $\{\bfY_{k-\tau},\dots,\bfY_k\}$.
\item[\(\{\bfY_{k}^{(n)}\}_k\):] Vector representing the BAU-discretized time series for the $i$-th independent replicate.
\item[\(\bfSigma_{k+h}^{(n)}(\bfY_k^{(n)}; \bms{\alpha})\):] The covariance matrix associated with \(\bfY_{k+h}^{(n)}\).
\item[\(\mathcal{T}^G_{\tau} \equiv \{0, \delta, \dots, \tau \delta\}\):] 
The temporal discretization of the interval $[0, \tau\delta]$ into $\tau+1$ equally spaced points with step size $\delta$.
\item[\(\mathcal{L}(\bftheta, \boldsymbol{\alpha})\):] 
The likelihood function of the parameters \(\bftheta\) and \(\boldsymbol{\alpha}\), where \(\boldsymbol{\alpha}\) is the vector of parameters that describe the spatial dependence of the innovation spatial field.

\end{description}

\bibliographystyle{apalike}

\bibliography{ref}
\end{document}


\renewcommand{\thesection}{S.\arabic{section}}
\renewcommand{\thesubsection}{S.\arabic{section}.\arabic{subsection}}
\renewcommand{\thesubsubsection}{S.\arabic{section}.\arabic{subsection}.\arabic{subsubsection}}
\numberwithin{equation}{section}
\renewcommand{\theequation}{S.\arabic{section}.\arabic{equation}}
\renewcommand{\thefigure}{S.\arabic{figure}}
\renewcommand{\thetable}{S.\arabic{table}}


\maketitle


Here we provide additional details about the methodology (Green's function),  the discretization, the computation (Fast Fourier Transform), and the simulation experiments (performance metrics, extra figures).

\section{Further technical details}
\subsection{ Derivation of the Green's function in Equation (2) in the main text}

Let the Fourier transform of $\mathcal{C}(\bfs,t)$ be $\mathcal{F}[\mathcal{C}]$, where
\[
\mathcal{F}[\mathcal{C}](\bfkappa,t) \equiv \hat{\mathcal{C}}(\bfkappa,t) = \int_{\mathbb{R}^d} e^{-\iota \bfkappa^\top \bfs}\, \mathcal{C}(\bfs,t)\, \mathrm{d}\bfs,
\]
with inverse
\[
\mathcal{F}^{-1}[\hat{\mathcal{C}}](\bfs,t) \equiv \mathcal{C}(\bfs,t) = \dfrac{1}{(2\pi)^d} \int_{\mathbb{R}^d} e^{\iota \bfkappa^\top \bfs}\, \hat{\mathcal{C}}(\bfkappa,t)\, \mathrm{d}\bfkappa,
\]
where $\iota$ is the imaginary number in  the complex plane, and $\bfkappa = (\kappa_1, \dots, \kappa_d)^\top$. Taking the Fourier transform of Equation (1) in the main text and, using linearity,
\begin{equation}
\dfrac{\partial }{\partial t} \mathcal{F}[\mathcal{C}](\bfkappa,t) 
+ \gamma_1 \sum_{i=1}^{d} \mathcal{F}\!\Big[\dfrac{\partial\mathcal{C}}{\partial s_i}\Big]
- \gamma_2 \sum_{i=1}^{d} \mathcal{F}\!\Big[\dfrac{\partial^2 \mathcal{C}}{\partial s_i^2}\Big]
= 0, \quad \text{for } (\bfkappa,t)\in\mathbb{R}^d\times[t_k,t_k+\tau].
\label{eq:fourier_advection}
\end{equation}
For each $i=1,\dots,d$,
\[
\mathcal{F}\!\Big[\dfrac{\partial\mathcal{C}}{\partial s_i}\Big](\bfkappa,t) 
= \int_{\mathbb{R}^d} e^{-\iota \bfkappa^\top \bfs}\, \dfrac{\partial\mathcal{C}}{\partial s_i}(\bfs,t)\, \mathrm{d}\bfs.
\]
Integrating by parts with respect to $s_i$ we get, 
\[
\mathcal{F}\!\Big[\dfrac{\partial\mathcal{C}}{\partial s_i}\Big](\bfkappa,t) 
= \iota \kappa_i \hat{\mathcal{C}}(\bfkappa,t), \qquad
\mathcal{F}\!\Big[\dfrac{\partial^2\mathcal{C}}{\partial s_i^2}\Big](\bfkappa,t) 
= -\kappa_i^2 \hat{\mathcal{C}}(\bfkappa,t).
\]
Note that
the integration-by-parts step uses the fact that a boundary term vanishes as $|s_i| \to \infty$.
This holds under the mild assumption that, for each fixed $t$, 
$\mathcal{C}(\cdot,t)$ and $(\partial \mathcal{C}/\partial s_i)(\cdot,t)$ belong to $L^1(\mathbb{R}^d;\mathbb{R})$,
which in particular implies $\mathcal{C}(\bfs,t) \to 0$ as $\|\bfs\| \to \infty$ \citep{brezis2011functional,evans2010pde}. \\
Substituting these expressions into Equation \eqref{eq:fourier_advection}, we obtain 
\[
\dfrac{\partial }{\partial t} \hat{\mathcal{C}}(\bfkappa,t)
+ \gamma_1 \sum_{i=1}^{d} \iota \kappa_i \hat{\mathcal{C}}(\bfkappa,t)
+ \gamma_2 \sum_{i=1}^{d} \kappa_i^2 \hat{\mathcal{C}}(\bfkappa,t) = 0,
\]
or equivalently,
\[
\dfrac{\partial \hat{\mathcal{C}}}{\partial t} (\bfkappa,t)
= -\big(\iota \gamma_1 \mathbf{1}^\top\bfkappa + \gamma_2 \|\bfkappa\|^2\big)\, \hat{\mathcal{C}}(\bfkappa,t),
\]
where $\mathbf{1}$ is the $d$-dimensional vector of 1s. We denote the equation above simply as $du/dt = -au$, where
$u(t) \equiv \hat{\mathcal{C}}(\bfkappa,t)$, and $a \equiv \iota \gamma_1 \mathbf{1}^\top\bfkappa + \gamma_2 \|\bfkappa\|^2$, for a fixed $\bfkappa$.
Then,
\[
u(t) \propto \, e^{-a t}.
\]
At the starting value $t=t_k$ in the interval $[t_k, t_k + \tau]$, we write $\hat{\mathcal{C}}(\bfkappa,t_k) \equiv \hat{\mathcal{C}}_0(\bfkappa)$. Hence 
\[
u(t_k+\tau) = e^{-a\tau}\, \hat{\mathcal{C}}_0(\bfkappa),
\]
or in terms of $\hat{\mathcal{C}}(\bfkappa,\cdot)$,
\[
\hat{\mathcal{C}}(\bfkappa,t_k+\tau)
= \exp\!\Big\{-(\iota \gamma_1 \mathbf{1}^\top\bfkappa + \gamma_2 \|\bfkappa\|^2)\tau\Big\}\, \hat{\mathcal{C}}_0(\bfkappa).
\]
Its inverse Fourier transform is,
\[
\mathcal{C}(\bfs,t_k + \tau)
= \dfrac{1}{(2\pi)^d}\! \int_{\mathbb{R}^d}\! 
e^{\iota \bfkappa^\top\bfs}\, 
\hat{\mathcal{C}}_0(\bfkappa)\,
\exp\!\Big\{-(\iota \gamma_1 \mathbf{1}^\top\bfkappa + \gamma_2 \|\bfkappa\|^2)\tau\Big\}\, \mathrm{d}\bfkappa.
\]
Using $\hat{\mathcal{C}}_0(\bfkappa) = \int_{\mathbb{R}^d} e^{-\iota \bfkappa^\top \bfu}\, \mathcal{C}(\bfu,t_k)\, \mathrm{d}\bfu$, and applying Fubini’s theorem, we have
\[
\mathcal{C}(\bfs,t_k+\tau)
= \int_{\mathbb{R}^d} 
\Bigg[\dfrac{1}{(2\pi)^d}\!\int_{\mathbb{R}^d}\!
\exp\!\Big\{\iota \bfkappa^\top (\bfs - \bfu) - \iota \gamma_1 \tau \mathbf{1}^\top\bfkappa - \gamma_2 \tau \|\bfkappa\|^2\Big\}
\, \mathrm{d}\bfkappa\Bigg] \mathcal{C}(\bfu,t_k)\, \mathrm{d}\bfu.
\]
Hence, for $\bfgamma = (\gamma_1, \gamma_2)^\top$, where $\gamma_2 > 0$, the Green's function is
\[
g(\mbf{r},\tau;\bfgamma)
= \dfrac{1}{(2\pi)^d} \int_{\mathbb{R}^d} 
\exp\!\Big\{\iota \bfkappa^\top\mbf{r} 
- \iota \gamma_1 \tau \mathbf{1}^\top\bfkappa
- \gamma_2 \tau \|\bfkappa\|^2\Big\} \mathrm{d}\bfkappa,
\]
which is real-valued and can be evaluated through a standard Gaussian integral to be: 
\[
g(\mbf{r},\tau;\bfgamma)
= \dfrac{1}{(4\pi \gamma_2 \tau)^{d/2}}
\exp\!\Big(-\dfrac{\|\mbf{r}-\tau \gamma_1 \mathbf{1}\|^2}{4\gamma_2\tau}\Big), \quad \tau>0.
\]
Finally, substituting $\tau=h\delta$ yields the desired result in Equation (2) of the main text.

\subsection{Further details on the discretization of the FNO input}

In the notation of Section~2.2 from the main text, the input field 
$Y_0^{(k)}(\cdot)$, for $k \in \{\tau+1, \dots, T-h\}$, evaluated on 
$\mathcal{D}^G \times \mathcal{T}^G$, where 
$\mathcal{T}^G \equiv \{0, \delta, \dots, \tau \delta\}$, is a tensor of size 
$m_1 \times \dots \times m_{d+1}$. 
Similarly, the output of layer $l$, $\mathbf{v}_l(\cdot)$, evaluated on 
$\mathcal{D}^G \times \mathcal{T}^G$, is a tensor of size 
$d_v \times m_1 \times \dots \times m_{d+1}$, for $l = 1, \ldots, L$. This tensor formulation lends itself to efficient computation on graphics and tensor processing units. In particular, the local operations $\mathcal{P}[\,\cdot\,], \mathcal{Q}[\,\cdot\,]$, and $\mathcal{W}_l[\,\cdot\,]$, for $l = 1,\dots,L$, simply reduce to multiple batch multiplications and additions that can be done in an embarrassingly parallel fashion for each point in $\mathcal{D}^G \times \mathcal{T}^G$. 

\subsection{Computation of discrete Fourier transforms}\label{sec:appendix_A}

The tensor formulation of Section 2.2 from the main text allows us to take advantage of the discrete Fourier transform (DFT). 
Computation of Equations (21) and (22) have complexity $\mathcal{O}((\prod_{j=1}^{d+1} \tilde{m}_j)^2)$, where $\tilde{m}_j$ are defined below Equation (22) in the main text. To speed up computation, in practice we use the Cooley–Tukey Fast Fourier Transform (FFT) algorithm \citep{tukey_fft}, where we let $\tilde{m}_j= 2^{q_j}, j = 1,\ldots , d+1$, and we compute the Fourier transforms at even and odd indices recursively. The computational complexity of the FFT is $\mathcal{O}((\prod_{j=1}^{d+1} \tilde{m}_j) \log(\prod_{j=1}^{d+1} \tilde{m}_j))$, which is exponentially faster than the naive DFT. Similarly, the Cooley-Tukey FFT algorithm can be used to compute the inverse DFT (IDFT), as defined in Equation (20), exponentially faster than the naive IDFT.
We use the \texttt{pytorch} software package to implement the multi-dimensional FFT for our applications.

\section{Further simulation-experiment details}
\subsection{Performance metrics}

We assess the forecasting performance of each model using metrics evaluated at $N_0$ test instances, namely the time series of spatial data excluded from training. We compute these metrics by comparing the forecasts from all the models with the true process values at the corresponding spatial locations.  The diagnostics we consider are the mean squared prediction error (MSPE), the prediction interval coverage probability (PICP) of a nominal 95\% prediction interval, and the mean prediction interval width (MPIW). These are defined as
\begin{align*}
    \text{MSPE} &= \dfrac{1}{N_0} {\dfrac{1}{|\mathcal{D}^G|}\dfrac{1}{T-h-\tau}\sum_{j=1}^{N_0}\sum_{\bfs\in \mathcal{D}^G} \sum_{k = \tau+1}^{T-h} (Y^{(j)}(\bfs,t_{k+h}) - \hat{Y}^{(j)}(\bfs,t_{k+h}))^2},\\
     \text{PICP} &= \dfrac{1}{N_0} \dfrac{1}{|\mathcal{D}^G|}\dfrac{1}{T-h-\tau}\sum_{j=1}^{N_0}\sum_{\bfs\in \mathcal{D}^G} \sum_{k = \tau+1}^{T-h} \mathbbm{1}\{{Y^{(j)}(\bfs,t_{k+h})\in [L^{(j)}(\bfs,t_{k+h}),U^{(j)}(\bfs,t_{k+h})]}\}, \ \\
     \text{MPIW} &= \dfrac{1}{N_0} \dfrac{1}{|\mathcal{D}^G|}\dfrac{1}{T-h-\tau}\sum_{j=1}^{N_0}\sum_{\bfs\in \mathcal{D}^G} \sum_{k = \tau+1}^{T-h} [U^{(j)}(\bfs,t_{k+h}) - L^{(j)}(\bfs,t_{k+h})],
   \end{align*}
   where $L^{(j)}(\bfs,t_{k+h})$ and $U^{(j)}(\bfs,t_{k+h})$ are the lower and upper forecast bounds of the 95\% prediction interval of $Y^{(j)}(\bfs,t+h)$.

\subsection{Additional testing instances for Section 3}
   This section presents supplementary figures corresponding to Section~3 of the main text, illustrating the results for two additional testing instances.

\begin{figure}[H]
    \centering
    
    \begin{subfigure}{0.75\textwidth}
        \centering
        \begin{tabular}{cc}
          FNO-DST-H & FNO-DST-NH \\
          \includegraphics[width=0.48\textwidth]{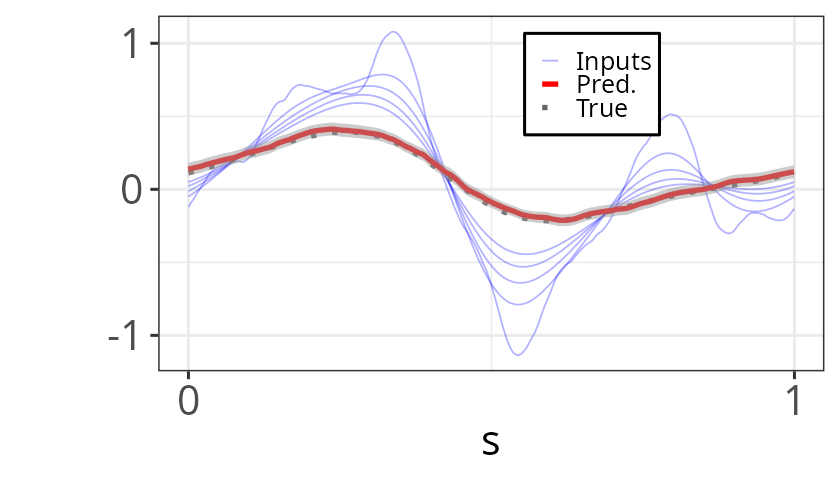} \vspace{6mm} & 
          \includegraphics[width=0.48\textwidth]{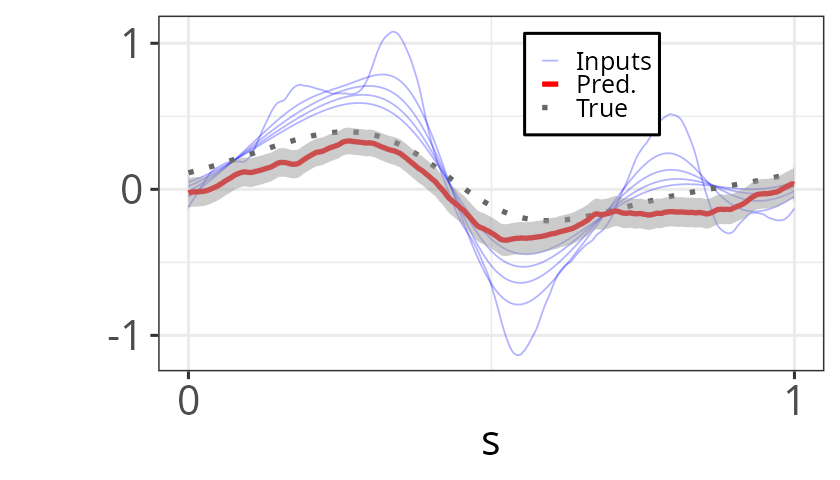} \vspace{6mm} \\
          \includegraphics[width=0.45\textwidth]{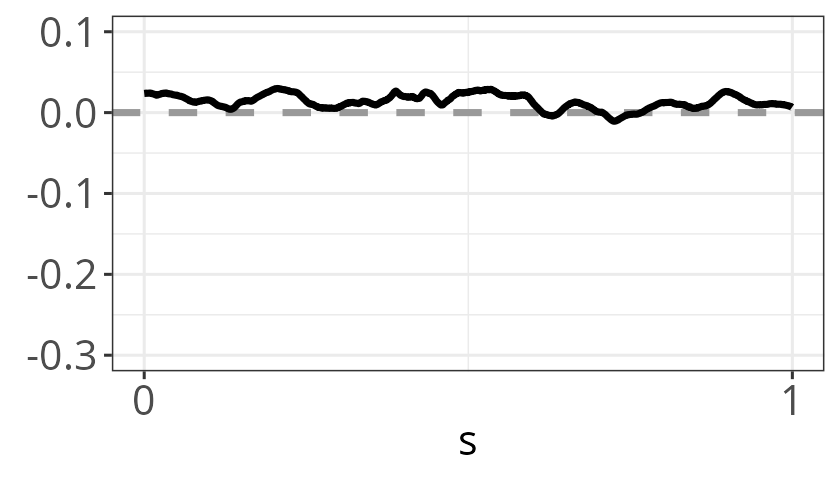} & 
          \includegraphics[width=0.45\textwidth]{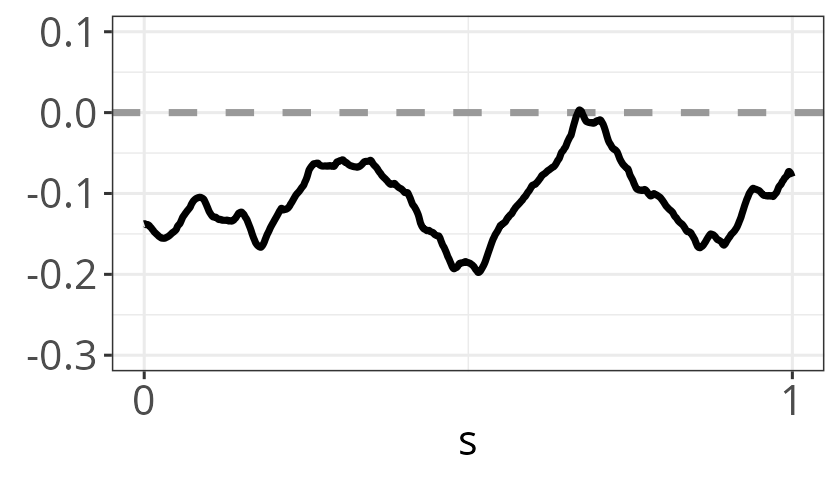} \\
        \end{tabular}
        \caption{Predicted fields (top row) and errors (bottom row) for random $\gamma = \gamma^{(1)}$.}
        \label{fig:burger-exp21}
    \end{subfigure}
    
    \vspace{0.5em} 
    
    \begin{subfigure}{0.75\textwidth}
        \centering
        \begin{tabular}{cc}
          FNO-DST-H & FNO-DST-NH \\
          \includegraphics[width=0.48\textwidth]{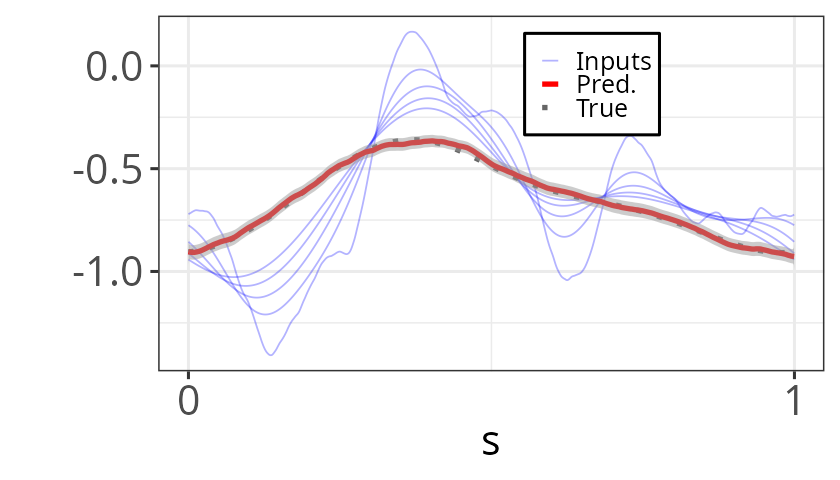} \vspace{6mm} & 
          \includegraphics[width=0.48\textwidth]{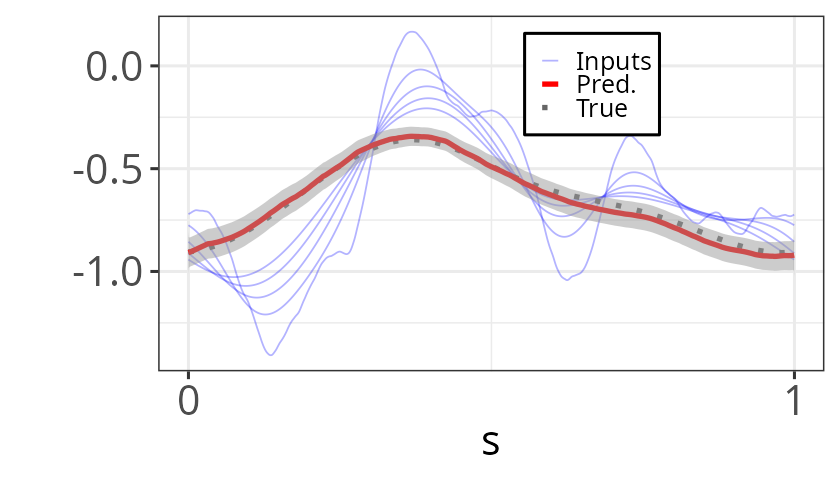} \vspace{6mm} \\
          \includegraphics[width=0.45\textwidth]{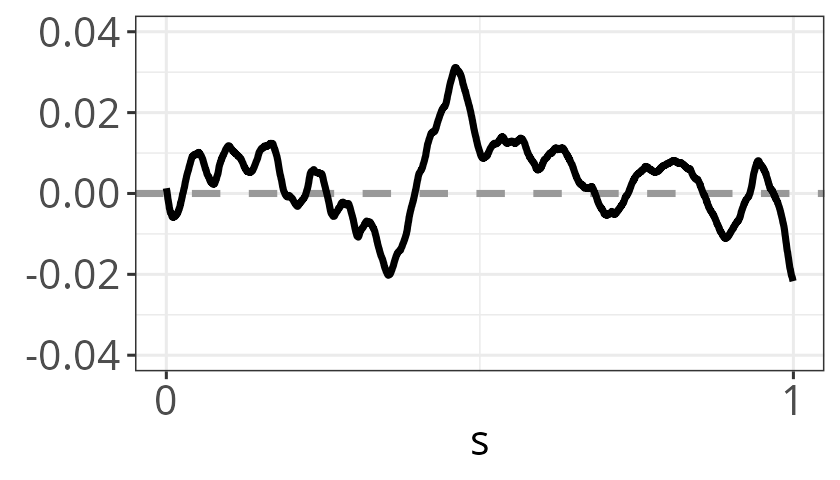} & 
          \includegraphics[width=0.45\textwidth]{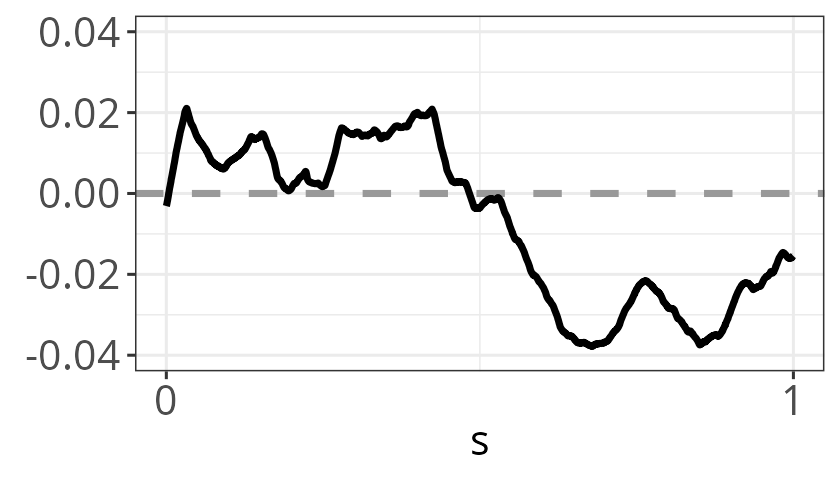} \\
        \end{tabular}
        \caption{Predicted fields (top row) and errors (bottom row) for fixed $\gamma = \gamma^{(2)}$.}
        \label{fig:burger-exp22}
    \end{subfigure}

    \caption{Second testing instance out of $N_0=50$ testing instances for, respectively, forecasts from FNO-DST-H and FNO-DST-NH models at $T = 10$. (a): $\gamma = \gamma^{(1)}$, (b): $\gamma = \gamma^{(2)}$. In the top panels of (a) and (b), the current and past four steps (blue lines), the true spatial field at $T=10$ (black dotted line), and the forecasts (red line) with 95\% prediction intervals (gray shaded area), are shown. }
    \label{fig:burger-both2}
\end{figure}
\begin{figure}[H]
    \centering
    \begin{tabular}{cccc}
        FNO-DST-H & ConvLSTM & STDK & Persistence \\

        \includegraphics[width=0.23\textwidth]{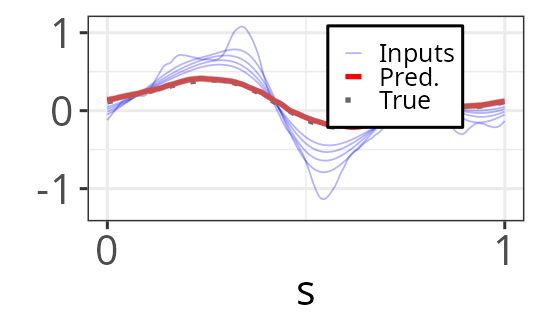} & 
        \includegraphics[width=0.23\textwidth]{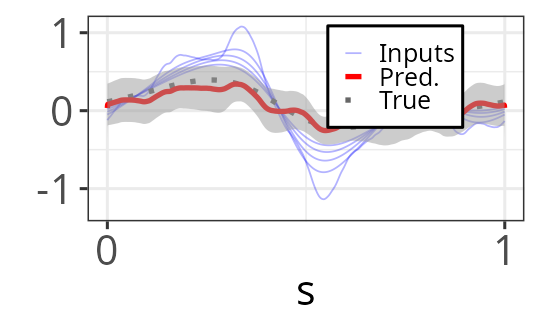} &
        \includegraphics[width=0.23\textwidth]{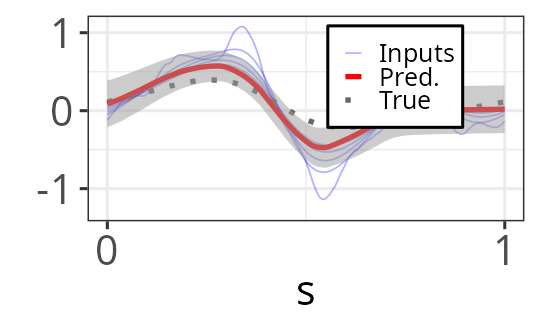} &
        \includegraphics[width=0.23\textwidth]{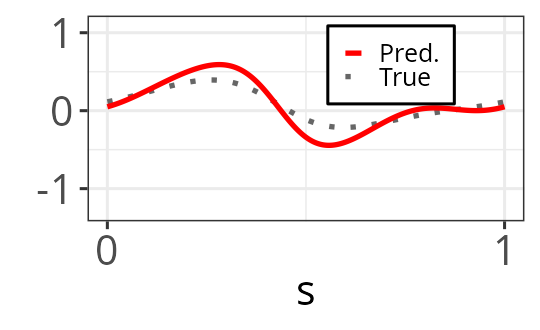} \\

        \includegraphics[width=0.23\textwidth]{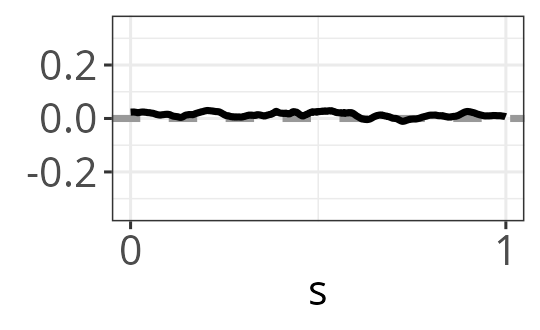} & 
        \includegraphics[width=0.23\textwidth]{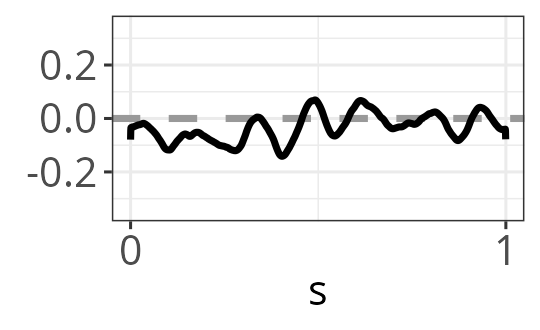} &
        \includegraphics[width=0.23\textwidth]{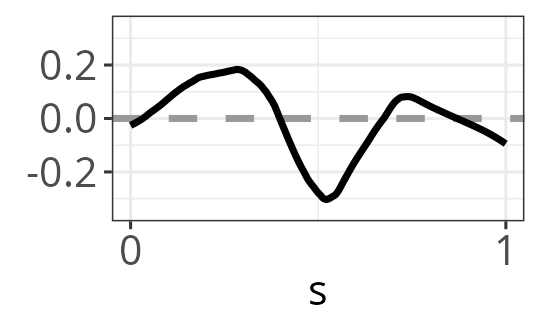} &
        \includegraphics[width=0.23\textwidth]{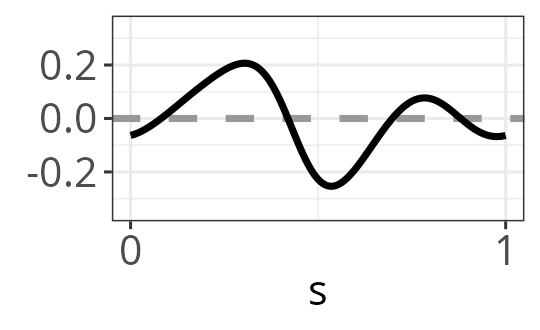} \\
    \end{tabular}
    \caption{Forecast fields (top row by model) and corresponding prediction errors (bottom row by model) for the second testing instance out of $N_0=50$ testing instances. This visualization illustrates both the forecast quality at $T=10$ and error distribution across space for the simulation setting with random $\gamma = \gamma^{(1)}$. In the top panels, the current and past four steps (blue lines), the true spatial field at $T=10$ (black dotted line), and the forecasts (red line) with 95\% prediction intervals (gray shaded area), are shown.  }
    \label{fig:burger2}
\end{figure}

\begin{figure}[H]
    \centering
    
    \begin{subfigure}{0.75\textwidth}
        \centering
        \begin{tabular}{cc}
          FNO-DST-H & FNO-DST-NH \\
          \includegraphics[width=0.48\textwidth]{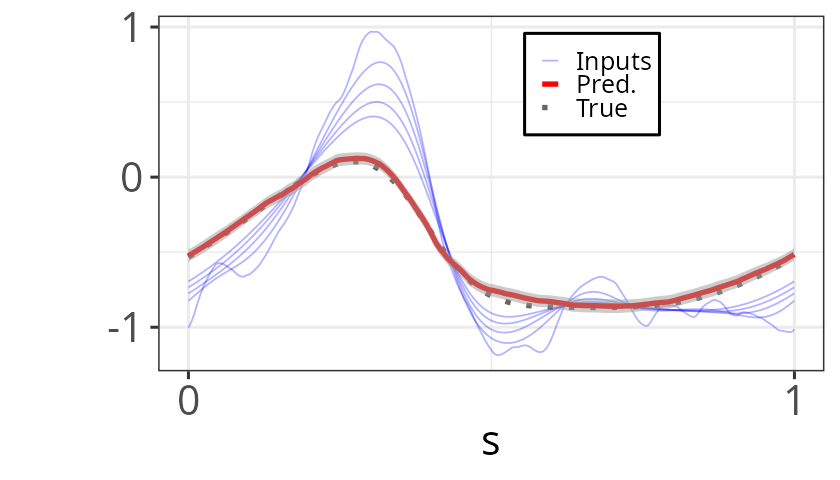} \vspace{6mm} & 
          \includegraphics[width=0.48\textwidth]{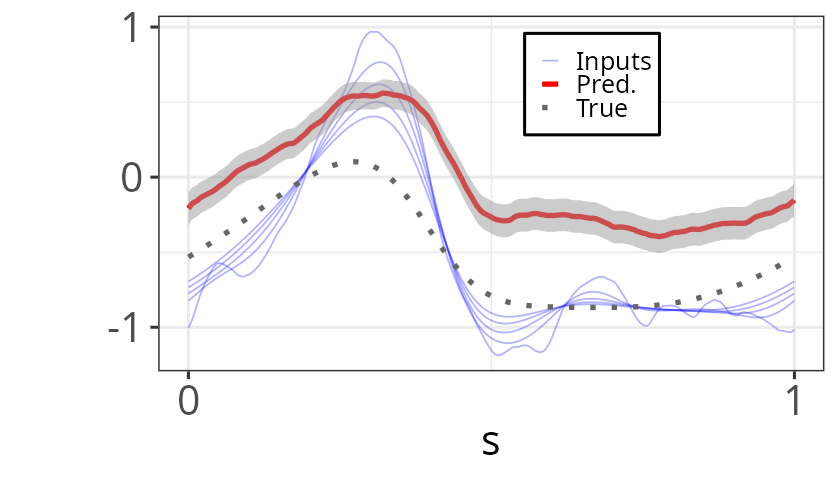} \vspace{6mm} \\
          \includegraphics[width=0.45\textwidth]{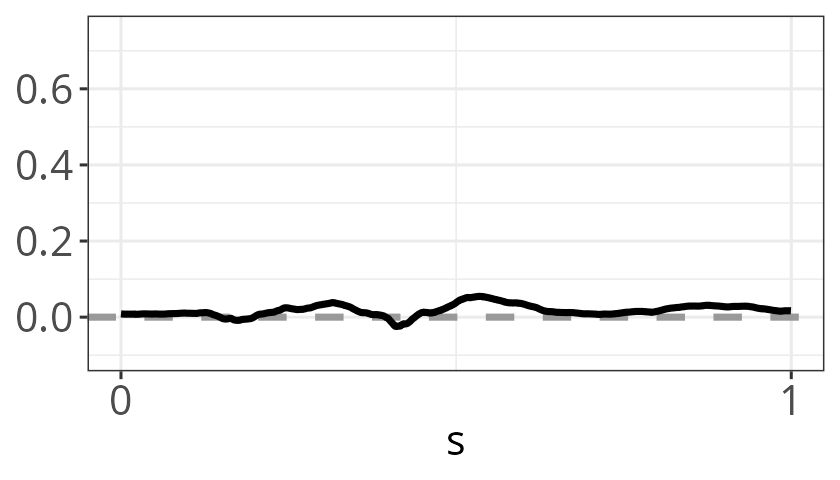} & 
          \includegraphics[width=0.45\textwidth]{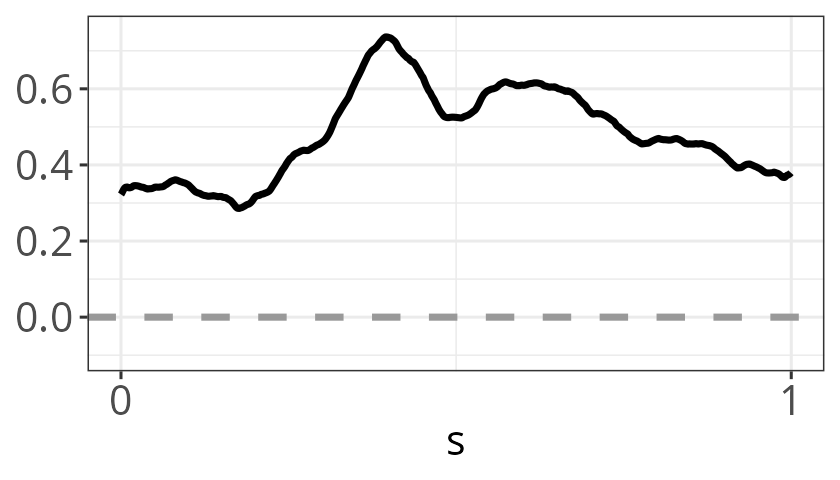} \\
        \end{tabular}
        \caption{Predicted fields (top row) and errors (bottom row) for random $\gamma = \gamma^{(1)}$.}
        \label{fig:burger-exp31}
    \end{subfigure}
    
    \vspace{0.5em} 
    
    \begin{subfigure}{0.75\textwidth}
        \centering
        \begin{tabular}{cc}
          FNO-DST-H & FNO-DST-NH \\
          \includegraphics[width=0.48\textwidth]{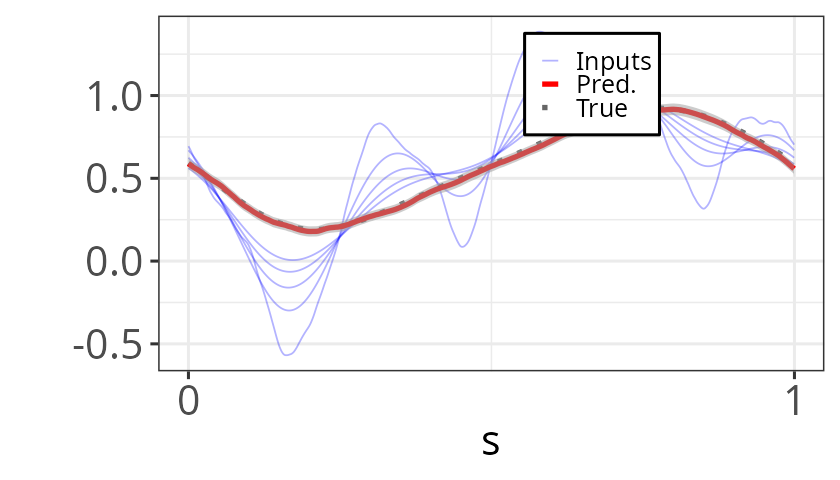}  \vspace{6mm} & 
          \includegraphics[width=0.48\textwidth]{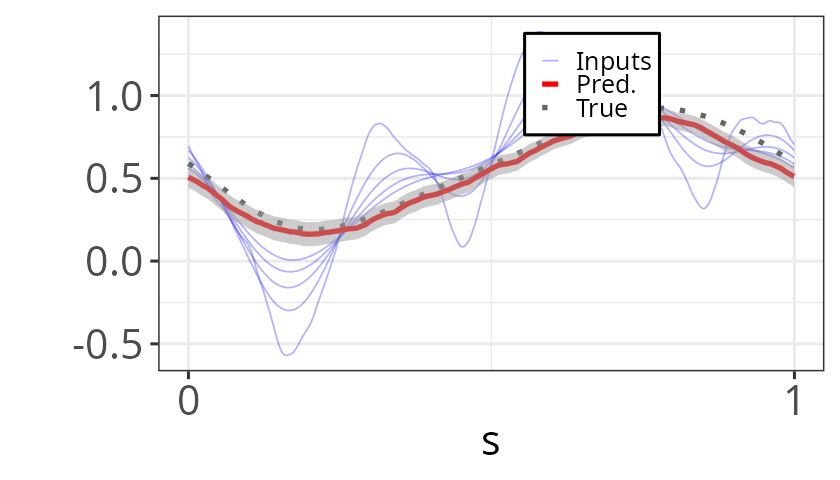} \vspace{6mm} \\
          \includegraphics[width=0.45\textwidth]{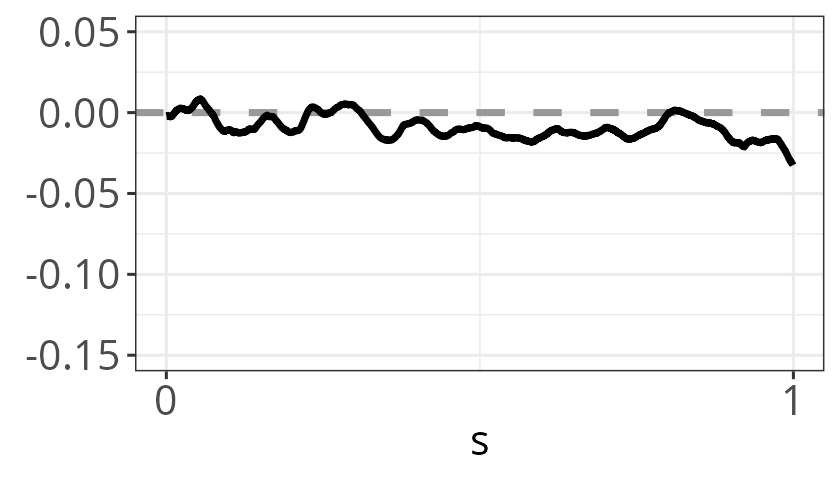} & 
          \includegraphics[width=0.45\textwidth]{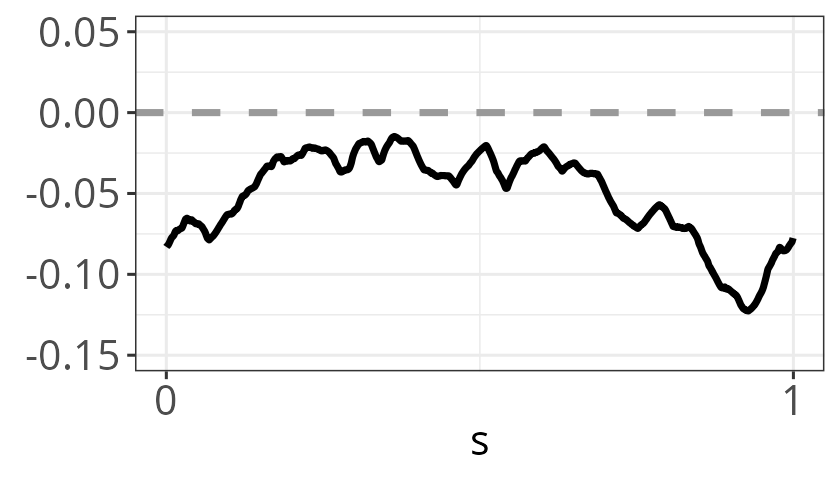} \\
        \end{tabular}
        \caption{Predicted fields (top row) and errors (bottom row) for fixed $\gamma = \gamma^{(2)}$.}
        \label{fig:burger-exp32}
    \end{subfigure}

    \caption{Third testing instance out of $N_0=50$ testing instances for, respectively, forecasts from FNO-DST-H and FNO-DST-NH models at $T = 10$. (a): $\gamma = \gamma^{(1)}$, (b): $\gamma = \gamma^{(2)}$. In the top panels of (a) and (b), the current and past four steps (blue lines), the true spatial field at $T=10$ (black dotted line), and the forecasts (red line) with 95\% prediction intervals (gray shaded area), are shown. }
    \label{fig:burger-both3}
\end{figure}

\begin{figure}[H]
    \centering
    \begin{tabular}{cccc}
        FNO-DST-H & ConvLSTM & STDK & Persistence \\

        \includegraphics[width=0.23\textwidth]{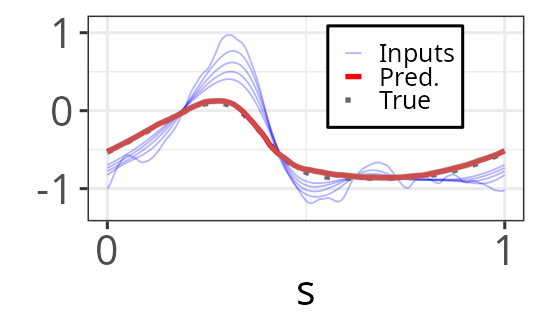} & 
        \includegraphics[width=0.23\textwidth]{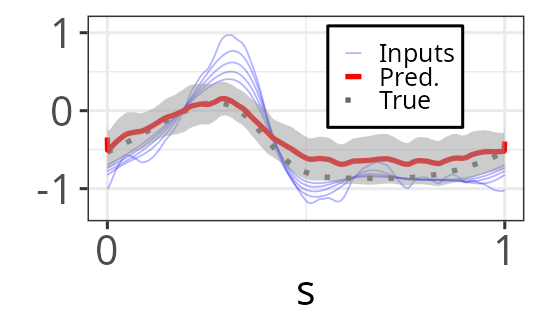} &
        \includegraphics[width=0.23\textwidth]{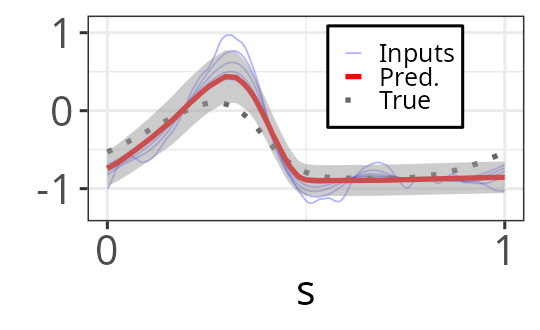} &
        \includegraphics[width=0.23\textwidth]{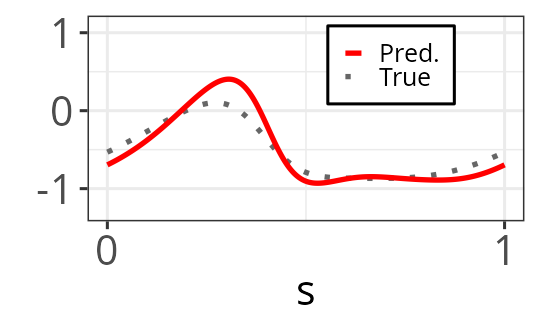} \\

        \includegraphics[width=0.23\textwidth]{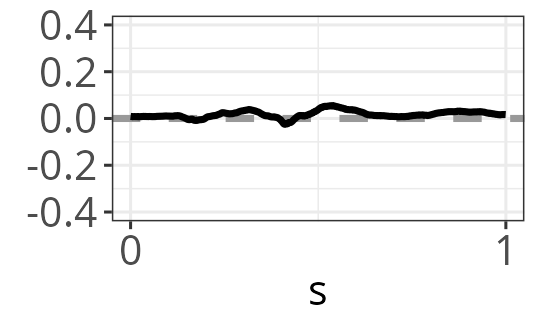} & 
        \includegraphics[width=0.23\textwidth]{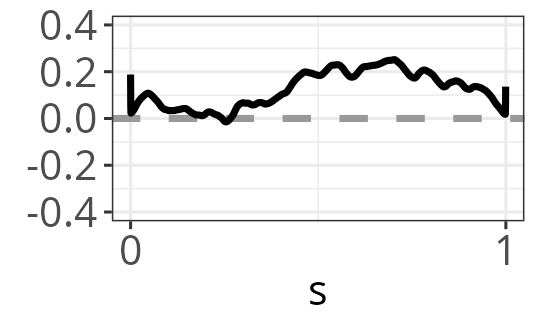} &
        \includegraphics[width=0.23\textwidth]{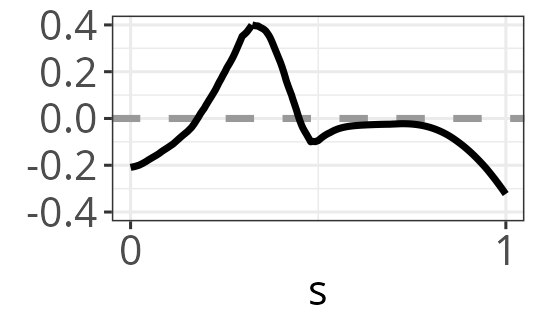} &
        \includegraphics[width=0.23\textwidth]{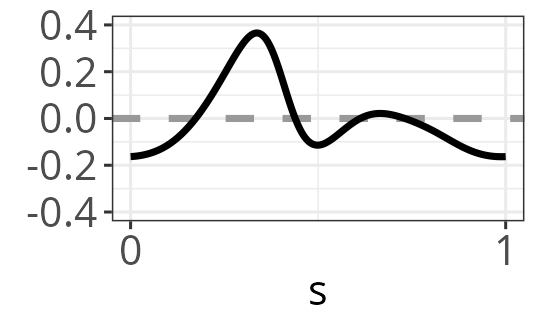} \\
    \end{tabular}
    \caption{Forecast fields (top row by model) and corresponding prediction errors (bottom row by model) for the third testing instance out of $N_0=50$ testing instances. This visualization illustrates both the forecast quality at $T=10$ and error distribution across space for the simulation setting with random $\gamma = \gamma^{(1)}$. In the top panels, the current and past four steps (blue lines), the true spatial field at $T=10$ (black dotted line), and the forecasts (red line) with 95\% prediction intervals (gray shaded area), are shown.  }
    \label{fig:burger3}
\end{figure}

\bibliographystyle{apalike}

\bibliography{ref}